\pdftrailerid{}          
\documentclass[twocolumn]{aastex631}
\newcommand*{\mytitle}{%
  FORECASTOR -- I. Finding Optics Requirements and Exposure times for the Cosmological
  Advanced Survey Telescope for Optical and UV Research mission
}
\newcommand*{\myshorttitle}{%
  FORECASTOR -- I. Photometric Exposure Time Calculator \& Examples
}
\newcommand*{\myshortauthors}{Cheng et al.}
\shorttitle{\myshorttitle}
\shortauthors{\myshortauthors}
\usepackage{formatting}
\let\oldAA\AA
\renewcommand*{\AA}{\ensuremath{\,\text{\oldAA}}\xspace}  
\newcommand*{\sn}{\ensuremath{\mathrm{S/N}}\xspace}  
\newcommand*{\cetc}{\textsc{castor\_etc}\xspace}  
\newcommand*{\asec}{\ensuremath{\,\mathrm{arcsec}}\xspace}  
\renewcommand*{\arcsec}{\ensuremath{''}\xspace}  
\renewcommand*{\farcs}{\ensuremath{\hspace{0.4ex}\mathclap{.}\hspace{-0.4ex}''\hspace{-0.25ex}}}  
\newcommand*{\sr}{\,\ensuremath{\mathrm{sr}}\xspace}  
\newcommand*{\erg}{\ensuremath{\,\mathrm{erg}}\xspace}  
\newcommand*{\px}{\ensuremath{\,\mathrm{px}}\xspace}  
\newcommand*{\e}{\ensuremath{\,\text{e\textsuperscript{-}}}\xspace}  
\newcommand*{\s}{\ensuremath{\,\mathrm{s}}\xspace}  
\newcommand*{\yr}{\ensuremath{\,\mathrm{yr}}\xspace}  
\newcommand*{\pow}[2][-1]{\ensuremath{\mathrm{#2}^{#1}}\xspace}  
\newcommand*{\degree}{\ensuremath{^\circ}\xspace}  
\renewcommand*{\deg}{\ensuremath{\,\mathrm{deg}}\xspace}  
\newcommand*{\Rsun}{\ensuremath{\,\mathrm{R}_\odot}\xspace}  
\newcommand*{\Msun}{\ensuremath{\,\mathrm{M}_\odot}\xspace}  
\newcommand*{\Hz}{\ensuremath{\,\mathrm{Hz}}\xspace}  
\newcommand*{\K}{\ensuremath{\,\mathrm{K}}\xspace}  
\newcommand*{\kpc}{\,\ensuremath{\mathrm{kpc}}\xspace}  
\newcommand*{\km}{\ensuremath{\,\mathrm{km}}\xspace}  
\newcommand*{\m}{\ensuremath{\,\mathrm{m}}\xspace}  
\newcommand*{\cm}{\ensuremath{\,\mathrm{cm}}\xspace}  
\newcommand*{\um}{\ensuremath{\,\text{\textmu m}}\xspace}  
\newcommand*{\nm}{\ensuremath{\,\mathrm{nm}}\xspace}  
\newcommand*{\oii}{[O\textsc{ii}]\xspace}  
\newcommand*{\npix}{\ensuremath{N_\mathrm{pix}}\xspace}  
\newcommand*{\nread}{\ensuremath{N_\mathrm{read}}\xspace}  
\newcommand*{\bpoisson}{\ensuremath{B_\mathrm{poisson}}\xspace}  
\newcommand*{\readnoise}{\ensuremath{R}\xspace}  
\newcommand*{\telescope}{\texttt{Telescope}\xspace}  
\newcommand*{\background}{\texttt{Background}\xspace}  
\newcommand*{\source}{\texttt{Source}\xspace}  
\newcommand*{\pointsource}{\texttt{PointSource}\xspace}  
\newcommand*{\extendedsource}{\texttt{ExtendedSource}\xspace}  
\newcommand*{\galaxysource}{\texttt{GalaxySource}\xspace}  
\newcommand*{\customsource}{\texttt{CustomSource}\xspace}  
\newcommand*{\photometry}{\texttt{Photometry}\xspace}  
\begin{document}

\title{\mytitle}

\correspondingauthor{Isaac Cheng}
\email{isaac.cheng@uwaterloo.ca, isaac.cheng.ca@gmail.com}

\author[0000-0002-8618-7990]{Isaac Cheng}
\affiliation{National Research Council of Canada, Herzberg Astronomy \& Astrophysics
Research Centre, 5071 West Saanich Road, Victoria, BC V9E 2E7, Canada}

\author[0000-0003-1428-5775]{Tyrone E.~Woods}
\affiliation{National Research Council of Canada, Herzberg Astronomy \& Astrophysics
Research Centre, 5071 West Saanich Road, Victoria, BC V9E 2E7, Canada}
\affiliation{Department of Physics and Astronomy, Allen Building, 30A Sifton Rd,
University of Manitoba, Winnipeg MB  R3T 2N2, Canada}

\author[0000-0003-1184-8114]{Patrick C\^ot\'e}
\affiliation{National Research Council of Canada, Herzberg Astronomy \& Astrophysics
Research Centre, 5071 West Saanich Road, Victoria, BC V9E 2E7, Canada}
\affiliation{Department of Physics and Astronomy, University of Victoria, Victoria, BC
V8W 2Y2, Canada}

\author[0009-0001-3879-3910]{Jennifer Glover}
\affiliation{Department of Physics and Astronomy, University of Victoria, Victoria, BC
V8W 2Y2, Canada}

\author[0009-0004-4161-3601]{Dhananjhay Bansal}
\affiliation{National Research Council of Canada, Herzberg Astronomy \& Astrophysics
Research Centre, 5071 West Saanich Road, Victoria, BC V9E 2E7, Canada}

\author[0009-0006-7454-3579]{Melissa Amenouche}
\affiliation{National Research Council of Canada, Herzberg Astronomy \& Astrophysics
Research Centre, 5071 West Saanich Road, Victoria, BC V9E 2E7, Canada}

\author[0000-0001-6434-7845]{Madeline A. Marshall}
\affiliation{National Research Council of Canada, Herzberg Astronomy \& Astrophysics
Research Centre, 5071 West Saanich Road, Victoria, BC V9E 2E7, Canada}

\author[0009-0001-9191-7222]{Laurie Amen}
\affiliation{National Research Council of Canada, Herzberg Astronomy \& Astrophysics
Research Centre, 5071 West Saanich Road, Victoria, BC V9E 2E7, Canada}

\author[0009-0006-4802-4601]{John Hutchings}
\affiliation{National Research Council of Canada, Herzberg Astronomy \& Astrophysics
Research Centre, 5071 West Saanich Road, Victoria, BC V9E 2E7, Canada}

\author[0000-0002-8224-1128]{Laura Ferrarese}
\affiliation{National Research Council of Canada, Herzberg Astronomy \& Astrophysics
Research Centre, 5071 West Saanich Road, Victoria, BC V9E 2E7, Canada}

\author[0000-0003-4134-2042]{Kim A. Venn}
\affiliation{Department of Physics and Astronomy, University of Victoria, Victoria, BC
V8W 2Y2, Canada}

\author[0000-0003-4849-9536]{Michael Balogh}
\affiliation{Department of Physics and Astronomy, University of Waterloo, Waterloo,
Ontario N2L 3G1, Canada}
\affiliation{Waterloo Centre for Astrophysics, University of Waterloo, Waterloo,
Ontario, N2L3G1, Canada}

\author[0000-0002-9632-1436]{Simon Blouin}
\affiliation{Department of Physics and Astronomy, University of Victoria, Victoria, BC
V8W 2Y2, Canada}

\author[0000-0001-5383-9393]{Ryan Cloutier}
\affiliation{Department of Physics \& Astronomy, McMaster University, 1280 Main St West,
Hamilton, ON, L8S 4L8, Canada}

\author[0000-0002-6865-2369]{Nolan Dickson}
\affiliation{Department of Astronomy and Physics, Saint Mary's University, 923 Robie
Street, Halifax, NS B3H 3C3, Canada}

\author[0000-0001-6217-8101]{Sarah Gallagher}
\affiliation{Department of Physics and Astronomy \& Institute of Earth and Space
Exploration, The University of Western Ontario, 1151 Richmond Street, London, ON N6A 3K7,
Canada}

\author{Martin Hellmich}
\affiliation{Department of Astronomy and Physics, Saint Mary's University, 923 Robie
Street, Halifax, NS B3H 3C3, Canada}

\author[0000-0003-2927-5465]{Vincent H\'enault-Brunet}
\affiliation{Department of Astronomy and Physics, Saint Mary's University, 923 Robie
Street, Halifax, NS B3H 3C3, Canada}

\author[0000-0002-0581-6506]{Viraja Khatu}
\affiliation{Department of Physics and Astronomy \& Institute of Earth and Space
Exploration, The University of Western Ontario, 1151 Richmond Street, London, ON N6A 3K7,
Canada}

\author[0000-0002-2958-0593]{Cameron Lawlor-Forsyth}
\affiliation{Department of Physics and Astronomy, University of Waterloo, Waterloo,
Ontario N2L 3G1, Canada}
\affiliation{Waterloo Centre for Astrophysics, University of Waterloo, Waterloo,
Ontario, N2L3G1, Canada}

\author[0009-0009-2522-3685]{Cameron Morgan}
\affiliation{Department of Physics and Astronomy, University of Waterloo, Waterloo,
Ontario N2L 3G1, Canada}
\affiliation{Waterloo Centre for Astrophysics, University of Waterloo, Waterloo,
Ontario, N2L3G1, Canada}

\author[0000-0001-9002-8178]{Harvey Richer}
\affiliation{Department of Physics and Astronomy, University of British Columbia,
Vancouver, BC V6T 1Z1, Canada}

\author[0000-0002-7712-7857]{Marcin Sawicki}
\affiliation{Department of Astronomy and Physics, Saint Mary's University, 923 Robie
Street, Halifax, NS B3H 3C3, Canada}

\author{Robert Sorba}
\affiliation{Department of Astronomy and Physics, Saint Mary's University, 923 Robie
Street, Halifax, NS B3H 3C3, Canada}

\received{2023 October 3}
\revised{2024 January 31}
\accepted{2024 February 12}
\published{2024 March 26}
\submitjournal{\aj}

\begin{abstract}

The Cosmological Advanced Survey Telescope for Optical and ultraviolet Research
(\textit{CASTOR}) is a proposed Canadian-led 1\m-class space telescope that will carry out
ultraviolet and blue-optical wide-field imaging, spectroscopy, and photometry. {\it
CASTOR} will provide an essential bridge in the post-Hubble era, preventing a protracted
UV-optical gap in space astronomy and enabling an enormous range of discovery
opportunities from the solar system to the nature of the Cosmos, in conjunction with the
other great wide-field observatories of the next decade (e.g., \textit{Euclid},
\textit{Roman}, Vera Rubin). FORECASTOR (Finding Optics Requirements and Exposure times
for \textit{CASTOR}) will supply a coordinated suite of mission-planning tools that will
serve as the one-stop shop for proposal preparation, data reduction, and analysis for the
{\it CASTOR} mission. We present the first of these tools:~a pixel-based, user-friendly,
extensible, multi-mission exposure time calculator (ETC) built in Python, including a
modern browser-based graphical user interface that updates in real time. We then provide
several illustrative examples of FORECASTOR's use that advance the design of planned
legacy surveys for the {\it CASTOR} mission: a search for the most massive white dwarfs in
the Magellanic Clouds; a study of the frequency of flaring activity in M stars, their
distribution and impacts on habitability of exoplanets; mapping the proper motions of
faint stars in the Milky Way; wide and deep galaxy surveys; and time-domain studies of
active galactic nuclei.

\end{abstract}


\section{Introduction}\label{sec:intro}

Spanning a wavelength range from the hydrogen-ionizing threshold (912\AA) to roughly the
onset of human vision ($\sim$4000\AA), the ultraviolet sky is dominated by massive stars,
their remnants, and their light scattered by dust through the Milky Way; by interstellar
shocks; by rapidly-star forming galaxies and active galactic nuclei; and by accreting
black holes and other compact objects. This invaluable window is almost totally
inaccessible from beneath the Earth's atmosphere, necessitating observations from space.
The early 1970s marked the dawn of UV space astronomy, with experiments including Apollo
16's Far Ultraviolet Camera \& Spectrograph \citep{Apollo16FUV}, Thor-Delta-1A
\citep{TD1a}, and Orion 1 \& 2 \citep{Orion}.

These initial efforts paved the way for the enormous success of the International
Ultraviolet Explorer \citep[\textit{IUE}, 1978--1996;][]{IUE}, the Far Ultraviolet
Spectroscopic Explorer \citep[\textit{FUSE}, 1999--2007;][]{FUSE}, and the Galaxy
Evolution Explorer \citep[\textit{GALEX}, 2003--2013;][]{GALEX} which together provided
revolutionary insight into a host of astrophysical sources, from interstellar dust, to
powerful distant quasars, and to massive star formation across the Universe \citep[see
e.g.,][for reviews]{Linsky,GomezdeCastro}. Today, access to the ultraviolet is maintained
thanks to the Hubble Space Telescope \citep{HSTCOS, HST_STIS}, the Ultraviolet-Optical
Telescope (\textit{UVOT}) onboard the Neil Gehrels Swift Observatory \citep{Swift_UVOT},
the optical-UV monitor onboard XMM-Newton \citep{XMM_OM}, and the Ultraviolet Imaging
Telescope (\textit{UVIT}) onboard the \textit{AstroSat} mission \citep{UVIT}. However,
while the Decadal Survey on Astronomy \& Astrophysics
(Astro2020)\footnote{\url{https://nap.nationalacademies.org/catalog/26141/pathways-to-discovery-in-astronomy-and-astrophysics-for-the-2020s}}
has strongly endorsed a next-generation infrared/optical/UV space telescope \citep[see
also][]{LUVOIR}, any planned launch awaits the 2040s, threatening a protracted gap in UV
astronomical capability in the coming years.

The Cosmological Advanced Survey Telescope for Optical and ultraviolet Research
(\textit{CASTOR}) is a proposed mission, led by the National Research Council of Canada
and the Canadian Space Agency, to address this need by providing wide-field
($\sim$0.25$\deg^{2}$) UV and optical imaging at Hubble-like resolution ($\sim$0\farcs15).
This will enable an enormous improvement in discovery efficiency together with UV and blue
optical spectroscopy \citep{CASTOR}. With a planned launch in the late 2020s, {\it CASTOR}
will image $\sim$5\% of the sky within its first five years, reaching a u-band depth 1.3
magnitudes deeper than LSST \citep{CASTOR_SMS}, and provide the widest, deepest, and
highest-resolution legacy survey available in the UV and blue-optical. Also included in
the present {\it CASTOR} reference design are low-resolution (\textit{R}$\sim$300--420)
grism-mode spectroscopy in the UV- and u-bands over the entire field of view, and a
digital micro-mirror device (DMD)-based configurable UV multi-object spectrograph (UVMOS)
in a parallel field,  providing access to $\lambda\sim\text{1500--3000}\AA$ with
$\text{\textit{R}}\sim1500$.

\textit{CASTOR} will enable transformative science in virtually every subfield of
astronomy, from cosmology to black hole astrophysics to exoplanet atmospheres to mapping
the outer solar system. At the same time, it will provide an enormous legacy of archival
observations strongly complementary to other wide-field surveys over the next decade
(e.g., \textit{Euclid}, \textit{Roman}). Through a dedicated transient survey and a Target
of Opportunity (ToO) program, {\it CASTOR} will also provide a powerful tool for the study
of cosmic transients. The mission has recently completed its phase 0 study, and
readily-adaptable tools for assessing mission parameters and planning observations are
thus needed by the international {\it CASTOR} community.

Here, we present the first such tool developed for the {\bf F}inding {\bf  O}ptics {\bf
R}equirements and {\bf E}xposure times for {\bf CASTOR} (FORECASTOR) project:~a dedicated
pixel-based photometric exposure time calculator and associated web-based graphical user
interface. In Section \ref{sec:design}, we outline the present reference design for the
CASTOR mission. In Section \ref{sec:etc}, we describe the software implementation for our
exposure time calculator and describe the essentials of its functionality. Finally, in
Section \ref{sec:science_cases}, we use the calculator to develop a selection of several
illustrative science cases:~a highly complete survey for the most massive white dwarfs in
both Galactic and Magellanic globular clusters; the distribution and frequency of M star
flares and implications for habitability; highly-precise measurements of the proper
motions of faint objects; comprehensive wide and deep galaxy surveys; and time-domain
studies of active galactic nuclei.

\section{CASTOR Reference Design}\label{sec:design}

\begin{figure}[b]
    \centering
    \includegraphics[width=\columnwidth]{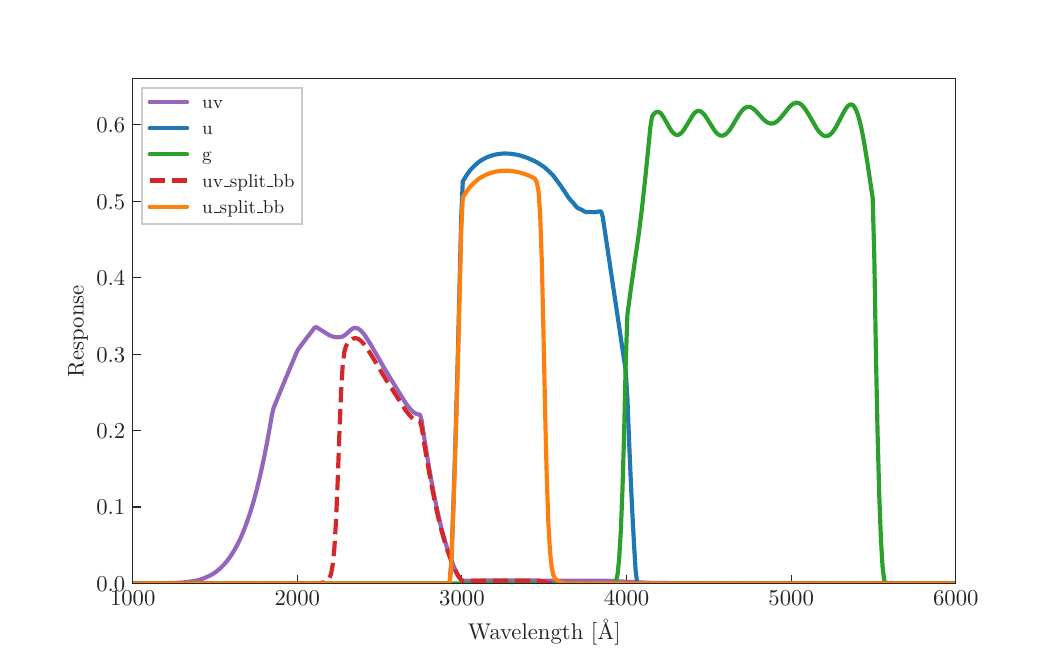}
    \caption{Filter response curves for the three baseline {\it CASTOR} filters (UV, u, g)
    and the response functions resulting from a proposed broadband filter that splits the
    UV- and u-bands.}
    \label{fig:filtercurves}
\end{figure}

CASTOR's reference optical design is a 1 metre unobscured three-mirror anastigmat
telescope. These mirrors divide the incident flux between three filters corresponding to
the NUV\nobreakdash-, u-, and g-bands, together providing simultaneous coverage over the
wavelength range $\sim$150--550\nm. Note that a filter wheel including a grism is
currently planned which will intercept the NUV- and u-channels, which may also include an
additional filter that will split the UV- and u-bands (see Figure \ref{fig:filtercurves}
and discussion in Section \ref{galaxies}). UV multi-object spectroscopy (UVMOS) will be
available in a parallel field \citep{UVMOS}. This design is presently being updated to
include a coated secondary folding mirror that will minimize ``red leak''---the detection
of long-wavelength photons in a blue filter due to imperfect blocking of optical and IR
light---in the NUV- and u-bands. This correction is reflected in the transmission curves
shown in Figure \ref{fig:filtercurves}.

The target point spread function (PSF) for photometry is a full-width half-maximum (FWHM)
of $\sim$0\farcs15 over the entire instantaneous field of view ($0.48\degree \times
0.50\degree \sim 0.24 ~\rm{sq.~deg.}$). The baseline detector design is for three 310
megapixel CMOS cameras with 10\um (0\farcs1) pixels, with stringent requirements for
minimal dark current at mission launch ($\sim$$10^{-4}\e\pow{\s}\pow{\px}$) and read noise
(3.0$\e\pow{\px}$). Taking all the above specifications together, the target sensitivity
is a limiting magnitude of \(\mathrm{AB}\!\sim\!27\) (at \(\sn\sim5\)) across all three
channels in $\sim$600\s (approximately 1/10th of an orbit). To confirm this, in the
following section we outline our exposure time calculator and its underlying assumptions.

\section{An Open-Source Exposure Time Calculator}\label{sec:etc}

The core of FORECASTOR's photometry ETC,
\cetc\footnote{\url{https://github.com/CASTOR-telescope/ETC}}, is written entirely in
Python with a strong emphasis on modular design and user-adaptability. We have endeavoured
to make this package as flexible and as maintainable as possible while providing a simple
user experience for those who do not need all of the possible customizations. All of the
critical components of the software are organized into distinct object classes, such that
a user may separately define instances of the telescope, background, and source, then pass
these to a photometry object which returns the desired signal-to-noise calculation. In the
following, we describe each of these classes in turn and the general software execution
flow. Appendix \ref{appdx:code_examples} contains a more detailed breakdown showing how to
use the code in practice.

\subsection{Telescope}\label{sec:telescope}

\newcommand*{\photzptsDescription}{%
  \IfFileExists{mtpro2.sty}{%
    \makecell{Photometric\\[4pt]Zero-Points (AB mag)}%
  }{%
    \makecell{Photometric\\Zero-Points (AB mag)}%
  }%
}
\newcommand*{\telescopehrule}{\noalign{\moveright 0.02\columnwidth \vbox{\hrule height
.15pt width 0.945\columnwidth}}}  

\begin{deluxetable}{cc}[htbp]
  \tablecaption{Some default \telescope parameters.\label{tab:telescope_params}}
  %
  %
  \tablehead{\colhead{\hspace*{4.5em}Parameter\hspace*{4.5em}} &
  \colhead{\hspace*{4.5em}Value\hspace*{4.5em}}}

  \startdata
  Passbands & [``uv'', ``u'', ``g''] \\
  \telescopehrule
  \multirow{3}{*}{Passband Limits (nm)} & ``uv'': [150, 300] \\
                                        & ``u'': [300, 400] \\
                                        & ``g'': [400, 550] \\
  \telescopehrule
  \multirow{3}{*}{\photzptsDescription} & ``uv'': 24.479 \\
                                                                     & ``u'': 24.564 \\
                                                                     & ``g'': 24.788 \\
  \telescopehrule
  \multirow{3}{*}{Pivot Wavelengths\tablenotemark{a} (nm)} & ``uv'': 226 \\
                                                           & ``u'': 345 \\
                                                           & ``g'': 475 \\
  \telescopehrule
  \multirow{3}{*}{Red Leak Thresholds\tablenotemark{b} (nm)} & ``uv'': 301 \\
                                                             & ``u'': 416 \\
                                                             & ``g'': 560 \\
  \telescopehrule
  \multirow{3}{*}{\(R(\text{passband}) \equiv \dfrac{A(\text{passband})}{E(B-V)}\)\tablenotemark{c}} & ``uv'': 7.06 \\
                                                                                                     & ``u'': 4.35 \\
                                                                                                     & ``g'': 3.31 \\
  \telescopehrule
  Pixel Scale (arcsec\pow{\px}) & 0.1 \\
  \telescopehrule
  FWHM of PSF\tablenotemark{d} (arcsec) & 0.15 \\
  \telescopehrule
  PSF Supersampling Factor\tablenotemark{e} & 20 \\
  \telescopehrule
  Dark Current\tablenotemark{f} (e\textsuperscript{-}\pow{\s}\pow{\px}) & \(10^{-4}\) \\
  \telescopehrule
  Read Noise (e\textsuperscript{-}\pow{\px}) & 3.0 \\
  \telescopehrule
  Mirror Diameter\tablenotemark{g} (cm) & 100 \\
  \telescopehrule
  Instantaneous Field of View\tablenotemark{g} (\degree) & \(0.48 \times 0.50\) \\
  \telescopehrule
  Detector Megapixel Count\tablenotemark{g} & 930 \\
  \telescopehrule
  Gain\tablenotemark{g} (e\textsuperscript{-}\,\pow{ADU}) & 2.0 \\
  \telescopehrule
  Bias\tablenotemark{g} (e\textsuperscript{-}) & 100 \\
  \enddata

  \tablenotetext{a}{Using the equal-energy (EE) convention given in Eq.~(A11) of
  \citet{TV05}.}
  \tablenotetext{b}{Flux longward of this wavelength is considered to be red leak for the
  given passband.}
  \tablenotetext{c}{Estimates taken from Table 2, column 3, rows ``NUV'', ``u'', and ``g''
  of \citet{yuan2013}.}
  \tablenotetext{d}{By default, we choose the FWHM value to be the largest estimate
  amongst the three passband PSFs. The FWHM of each PSF is estimated to be 0\farcs08,
  0\farcs12, and 0\farcs15 for the UV-, u-, and g-bands, respectively. Note that this FWHM
  value is only used for estimating an ``optimal aperture'' size, as discussed in Section
  \ref{sec:photometry}, and this value is easily modified by the user. Instead, we use the
  2D PSFs for actually generating the mock signals pixel-by-pixel.}
  \tablenotetext{e}{The oversampling factor is relative to the telescope's pixel scale,
  meaning our PSF files have pixels with side lengths of \(0\farcs1/20=0\farcs005\).}
  \tablenotetext{f}{Valid at detector's beginning-of-life and will increase linearly with
  time due to trapped proton radiation, reaching \(0.01\e\s^{-1}\px^{-1}\) by the end of
  five years.}
  \tablenotetext{g}{Currently not used for any photometry ETC calculations.}

  \vspace{5pt}
  \tablecomments{These values represent our current, most up-to-date knowledge of the
  telescope design and performance.}
\end{deluxetable}

The first step in using the \cetc Python package is always to define an instance of a
\telescope object. All aspects of the telescope are customizable. This includes, but is
not limited to, the number and name of passbands, passband response curves and the
passband limits, the PSF in each passband, pixel scale, read noise, the extinction
coefficients of each passband, etc. For convenience, each \telescope object is initialized
with sensible default values according to the most up-to-date information available at the
time.

The default parameters are maintained and updated in a central file; a summary of the most
critical of these is presented in Table \ref{tab:telescope_params}. Note, however, that
any of these values can be changed by the user, in which case one can pass in just the
modified value(s) to a particular \telescope instance rather than modifying the source
file directly.

Among the default parameters, \cetc includes the nominal reference design throughput
curves for the three \textit{CASTOR} photometric passbands:~UV, u, and g, shown in Figure
\ref{fig:filtercurves}. These, too, may be substituted for any arbitrary throughput
appropriate to a given detector, optics, and filter combination. Given some passband
response curve, the photometric zero-point and pivot wavelength are calculated
automatically. The photometric zero-point, defined as the AB magnitude which produces
\(1\e\pow{\s}\) in a given passband, is calculated assuming a flat spectrum in AB
magnitude and converges on the zero-point using either the secant or bisection method,
depending on the user's choice. The pivot wavelength is computed following Eq.~(A11) from
\cite{TV05} and uses Simpson's rule to approximate the integration.

\begin{figure*}[tbp]
  \centering
  \begin{minipage}[c]{0.32\textwidth}
    \vspace*{0pt}
    \includegraphics[width=\textwidth]{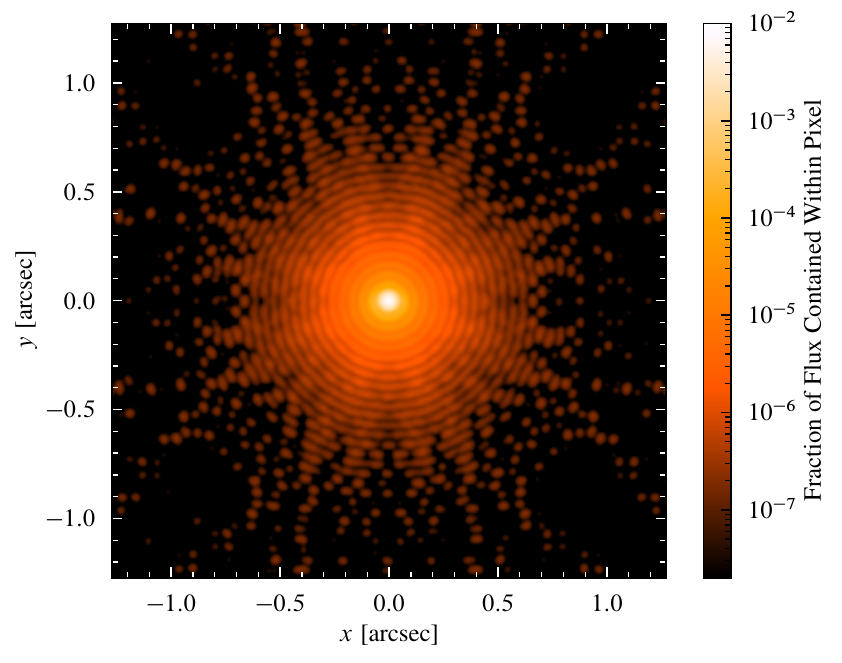}
  \end{minipage}
  \hfill
  \begin{minipage}[c]{0.32\textwidth}
    \vspace*{0pt}
    \includegraphics[width=\textwidth]{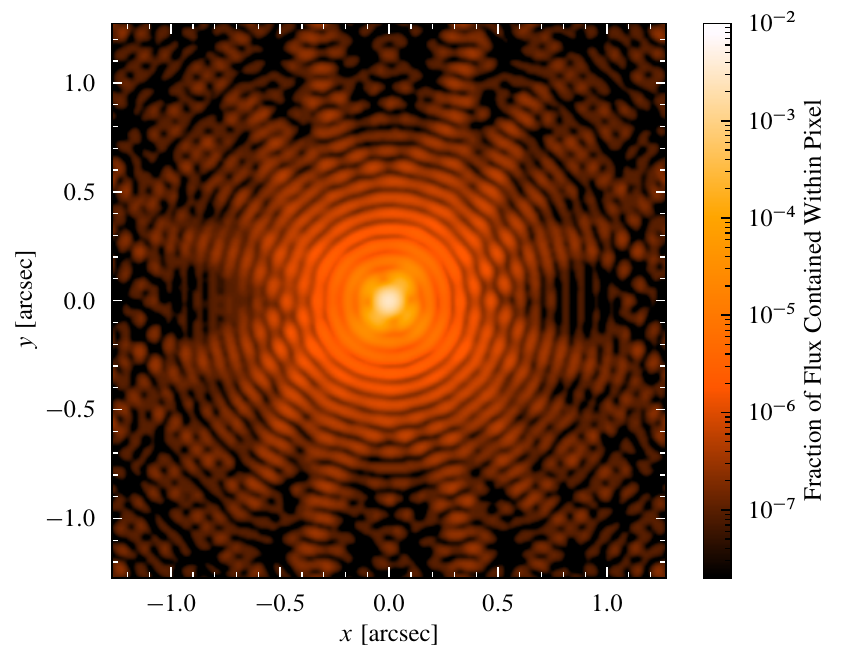}
  \end{minipage}
  \hfill
  \begin{minipage}[c]{0.32\textwidth}
    \vspace*{0pt}
    \includegraphics[width=\textwidth]{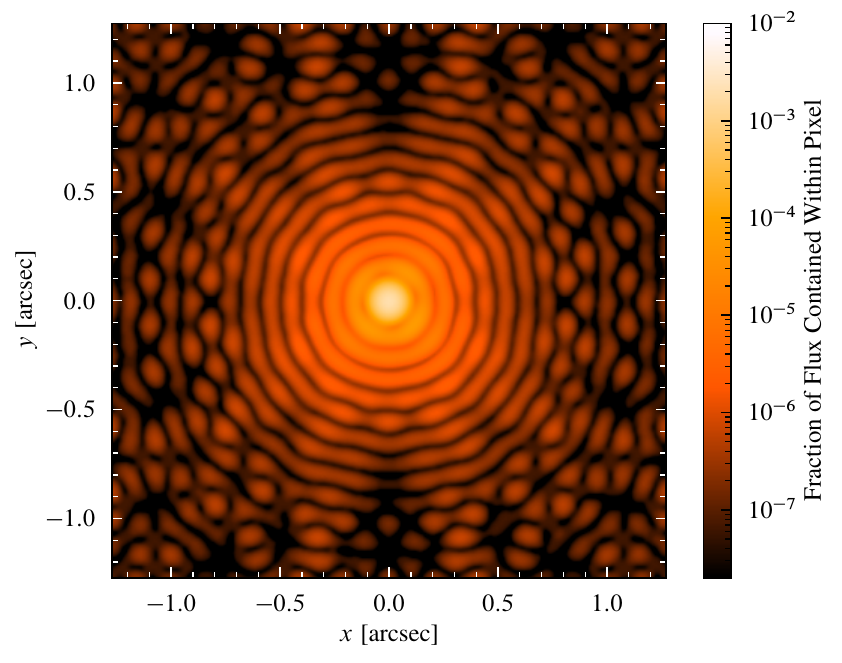}
  \end{minipage}
  \caption{Point spread functions for \textit{CASTOR}'s central field of view, in each of
  its UV- (left), u- (centre), and g-bands (right). The original PSF simulations are
  sampled at 10\(\times\) \textit{CASTOR}'s pixel scale, then interpolated to produce a
  grid which is 20\(\times\) the telescope's pixel scale.}
  \label{fig:psfs}
\end{figure*}

For each of \textit{CASTOR}'s passbands, \cetc also includes a default wavelength-averaged
PSF sampled at 20\(\times\) the telescope's pixel scale (i.e., each pixel has a side
length of \(0\farcs1/20=0\farcs005\)), as shown in Figure \ref{fig:psfs}. Supersampling
the PSF increases the accuracy of our calculations, as we explain in Section
\ref{sec:photometry}. Strictly speaking, these default PSFs are only valid near the centre
of \textit{CASTOR}'s field of view, but drop-in replacements to other PSFs are easily
accomplished through specifying the filepaths to the PSF files.

The ability to specify different PSFs and passbands that recompute zero-points and pivot
wavelengths as needed, together with the complete customizability discussed above, means
that theoretically, this package can be used for other missions as long as the telescope
detector is CCD- or CMOS-based or similar. A more thorough discussion and example showing
how to adapt FORECASTOR to other missions can be found in Appendix
\ref{appdx:adapting_forecastor}.

\subsection{Background}\label{sec:background}

The sky background is characterized by three parameters:~Earthshine, zodiacal light, and
geocoronal emission (i.e., ``airglow''). By default, the \background object uses spectra
from the Hubble Space Telescope \citep[as given by the][]{pysynphot} that provide
Earthshine and zodiacal light estimates \citep[also see Table 6.4 and the paragraph
preceding Figure~6.1 of][]{STISHandbook}. Users can also input their own Earthshine or
zodiacal light spectra. Alternatively, users can describe the background conditions by
specifying the sky background in AB magnitude per square arcsecond in each passband, which
will take precedence over any spectrum files. Note that these AB magnitudes should include
the effects of both Earthshine and zodiacal light.

At this point, the \background object does not have any geocoronal emission lines added to
it. Users can, however, add geocoronal emission lines of arbitrary wavelength, linewidth,
and flux. For convenience, we set the default airglow value to represent the \oii 2471\AA
emission line, which is centred at 2471\AA with a linewidth of 0.023\AA. \cetc provides
three predefined flux values (in units of \(\mathrm{erg}\cm^{-2}\s^{-1}\asec^{-2}\))
following those used by the HST STIS Handbook \citep[Table 6.5]{STISHandbook}:~``high''
($3.0\times 10^{-15}$), ``avg'' ($1.5\times10^{-15}$), and ``low'' ($1.5\times10^{-17}$).
By default, we assume a uniform sky background over the user-selected aperture (aperture
selection will be discussed in Section \ref{sec:photometry}), but spatially non-uniform
backgrounds may also be defined pixel-by-pixel.

\subsection{Source}\label{sec:source}

Once the \telescope and \background objects have been defined, the next step is to define
a \source object for mock observations.

In general, the creation of any \source object has three steps:
\begin{enumerate}
  \item \label{enum:source_types}Determining the type of the source (a point source, an
  extended source like a diffuse nebula, a galaxy, etc.).
  \item Describing the physical properties of the source, such as its spectrum (including
  any emission/absorption lines), redshift, distance, surface brightness profile (for
  extended sources or galaxies), etc.
  \item (Optional) Renormalizing the source spectrum. There are several normalization
  schemes available:~normalize to an AB magnitude within a passband, normalize to a total
  luminosity and distance, normalize a blackbody spectrum to a star at a given radius and
  distance. Note that these normalizations can be applied at any time (e.g., can be before
  or after the addition of spectral lines).
\end{enumerate}
There are several subclasses to aid with item \ref{enum:source_types}. Currently, we
have:~\pointsource, \extendedsource, and \galaxysource classes, which facilitate the
creation of mock point sources, diffuse extended sources, and galaxies, respectively. If
the user has a specific target in mind with a complicated spatial profile, the user can
upload a FITS file of the source to create a \customsource instance, and the data will be
automatically interpolated to the \telescope's pixel scale. This feature allows the user
to bypass the ETC's built-in source generation machinery while using the most of
FORECASTOR ETC's other functionality like sky background estimation and aperture
selection.

\cetc also provides utilities to generate the following spectra, which are all agnostic to
the \source class:~blackbody, power-law, emission line with different line shapes, and a
flat spectrum in either units of \(\mathrm{erg}\pow{\s}\pow[-2]{\cm}\pow{\text{\AA}}\),
\(\mathrm{erg}\pow{\s}\pow[-2]{\cm}\pow{\text{\Hz}}\), AB magnitude, or ST magnitude. For
convenience, we also provide stellar spectra from \citet{Pickles1998}, and spectra for
spiral and elliptical galaxies from \citet{Fioc1997}. Users can also load their own
spectra from data files to use in any of the \source subclasses. Finally, we support
adding emission and absorption lines with various line profiles to any spectrum, including
custom spectra.

\subsection{Photometry}\label{sec:photometry}

The \photometry class handles the final step in obtaining photometric estimates from
\cetc. We initialize the \photometry class by giving it our \telescope, \source, and
\background objects. Having a separate \photometry class allows for greater flexibility as
calculations are not tied to a specific \telescope, \background, or \source instance.

We use \textsc{photutils} \citep{photutils} to generate apertures with fractional pixel
contributions for our signal-to-noise (\sn) measurements, where the contribution of a
pixel is directly proportional to how much of the pixel is contained within the aperture.
Currently, we have support for rectangular apertures, elliptical apertures, and for point
sources only, ``optimal'' apertures.

The optimal aperture of a point source is a circular aperture centred on the source that
maximizes the \sn, assuming the only source of noise is shot noise (due to source counts,
sky background, dark current). By default, the optimal aperture's diameter is set to
1.4\(\times\) the telescope's FWHM, which is roughly the optimal aperture if the PSF is a
2D Gaussian. Since the PSFs are not Gaussians and differ between filters, however, the
true optimal aperture will be slightly different in each passband. Another factor that
causes the true optimal aperture to differ from this estimate is due to our photometry
calculation not only including Poisson noise, e.g., due to the sky background and dark
current (either of which may be set to be spatially varying), but also including noise due
to detector read outs, which is not modelled by a Poisson process. In fact, due to the
different source flux and sky background that passes through each filter, the relative
contribution of the non-Poissonian read noise in each band is different and thus the size
of the true optimal aperture varies from passband to passband even for the same PSF.
Therefore, while in the absence of any more refined estimate we recommend the default
factor of 1.4$\times$, we also allow the user to optionally specify their own custom
multiplicative factor to the telescope's FWHM when creating an optimal aperture.

In addition to the considerations above, \citet{Mighell1999} notes that brighter stars
also have larger optimal apertures than smaller stars. That being said, the \sn is fairly
insensitive to the exact size of the aperture when it is close to the optimal aperture
\citep{Pritchet1981}.

Once the aperture is specified, we are able to generate 2D maps of the signal and noise in
every pixel for each of the passbands. We create these 2D maps by convolving a noiseless
image at the PSFs' supersampled resolution with each passband's PSF, then binning down to
the telescope's resolution. For example, the default PSF is oversampled by a factor of 20,
meaning we bin blocks of \(20\times20=400\) subpixels down to 1 pixel. Thus, the higher
the sampling resolution of the PSF, the more accurate the calculations will be, as it
allows us to better approximate the continuous spatial distribution of flux.

We also determine the fraction of flux enclosed within the aperture (i.e., the encircled
energy) by comparing the flux within the aperture to some reference flux value. For point
sources, the reference flux is simply the sum of the PSF pixel values, which represents
100\% of the flux from the source. For galaxies and extended sources, the reference flux
is always defined using the noiseless image (i.e., before PSF convolution), since the
total amount of flux should be independent of the PSF. In particular, the reference flux
for galaxies is the flux contained within a centred elliptical aperture that is the same
size as the galaxy's half-light radius and of the same orientation. We then assume the
total flux from the galaxy to be twice this reference flux. For extended sources, we
assume 100\% of the flux is contained within the source's angular extent, which is true
for extended sources with a uniform surface brightness profile. Thus, the reference value
representing 100\% of the flux for an extended source is the signal obtained through a
centred elliptical aperture of the same dimensions and orientation as the extended source.

An enclosed flux fraction of 100\% corresponds to the magnitude of the source that the
user set. If the user normalizes a spectrum to, say, an AB magnitude of 25, then this AB
magnitude will be the AB magnitude of the source if 100\% of its flux was contained within
the aperture. If the user selects an aperture that only contains 50\% of the flux,
however, the effective AB magnitude will be dimmer.

These enclosed flux fractions are used to determine the number of electrons produced per
second on the detector in a given passband \(a\) using the following formula:
\begin{equation}\label{eq:counts}
  \text{e\textsuperscript{-}}\s^{-1}
  = 10^{-\frac{2}{5}\big[ m(a) + R(a) \times E(B-V) - Z(a) \big]} \times f(a),
\end{equation}
where \(m(a)\) is the apparent magnitude of the source through the passband (i.e.,
obtained through convolving the spectrum with the passband response curve), \(R(a)\) is
the extinction coefficient for that passband, \(E(B-V)\) is the reddening (which depends
on the telescope pointing), \(Z(a)\) is the photometric zero-point for that passband
(determined from the passband response curve), and \(f(a)\) is the fraction of flux
contained within the aperture.

The first two terms in the exponent of Eq.~\eqref{eq:counts} is the extinction-corrected
magnitude of the source. Thus, \(10^{-0.4[m(a)+R(a)\times E(B-V)]}\) converts the source
magnitude into flux. Scaling the extinction-corrected magnitude to the passband's
photometric zero-point (equivalent to dividing \(10^{-0.4[m(a)+R(a)\times E(B-V)]}\) by
\(10^{-0.4Z(a)}\)) converts the flux into the number of electrons produced per second on
the detector. Finally, multiplying by \(f(a)\) accounts for the fraction of flux enclosed
within the aperture; an aperture that encloses 50\% of the flux from a source will produce
half as much signal as an aperture that contains 100\% of the flux.

Note that the ETC does this calculation pixel-by-pixel, as \(f(a)\) is defined per pixel
and includes fractional pixel weighting, giving us a 2D array of the number of electrons
produced per second per pixel in the aperture.

Then, to calculate the signal-to-noise ratio \(\Sigma\) achieved in a given integration
time \(t\), we use the standard \sn formula \citep[see, e.g.,][]{WFC3_handbook}:
\begin{equation}\label{eq:snr_from_t}
  \Sigma(t) = \frac{Qt}{\sqrt{Qt+\npix\bpoisson t + \npix\nread\readnoise^2}},
\end{equation}
where \(Q\) is the total signal in the aperture (in units of
e\textsuperscript{-}\pow{\s}), \npix is the number of pixels in the aperture, \bpoisson is
the Poisson noise due to Earthshine, zodiacal light, geocoronal emission, and dark
current, \readnoise is the read noise, and \nread is the number of detector readouts. We
can clear the denominator and the square root in Eq.~\eqref{eq:snr_from_t}, and apply the
quadratic formula to solve for the integration time \(t\) needed to achieve a desired
signal-to-noise ratio \(\Sigma\):
\begin{align}
  t = &\Big[\Sigma^2 (Q + \npix\bpoisson) \nonumber \\
  &+ \sqrt{\Sigma^4 (Q + \npix\bpoisson)^2 + 4Q^2\Sigma^2\npix\nread\readnoise^2}\Big]
  \over
  \hfill2Q^2\hfill\makebox(0,0){\hspace{5pt}\raisebox{12pt}{\(.\)}}
\end{align}

To execute these calculations in practice, we use our 2D arrays describing the source
signal, sky background, and dark current of every pixel, accounting for fractional pixels.
The factor of \npix is implicitly included in these arrays, so simply summing these arrays
is enough to calculate the total signal, sky background, and dark current without further
multiplying by \npix. The only exception is the read noise and the number of detector
readouts, which are scalars and are multiplied by \npix rounded up to the nearest integer
(as you cannot read out a fraction of a pixel). We emphasize that this ETC is completely
pixel-based, so changing any value in the 2D arrays describing the source, background, or
dark current, or changing the number of effective \npix, will affect the photometry
calculations.

\begin{figure*}
  \centering
  \includegraphics[width=\textwidth]{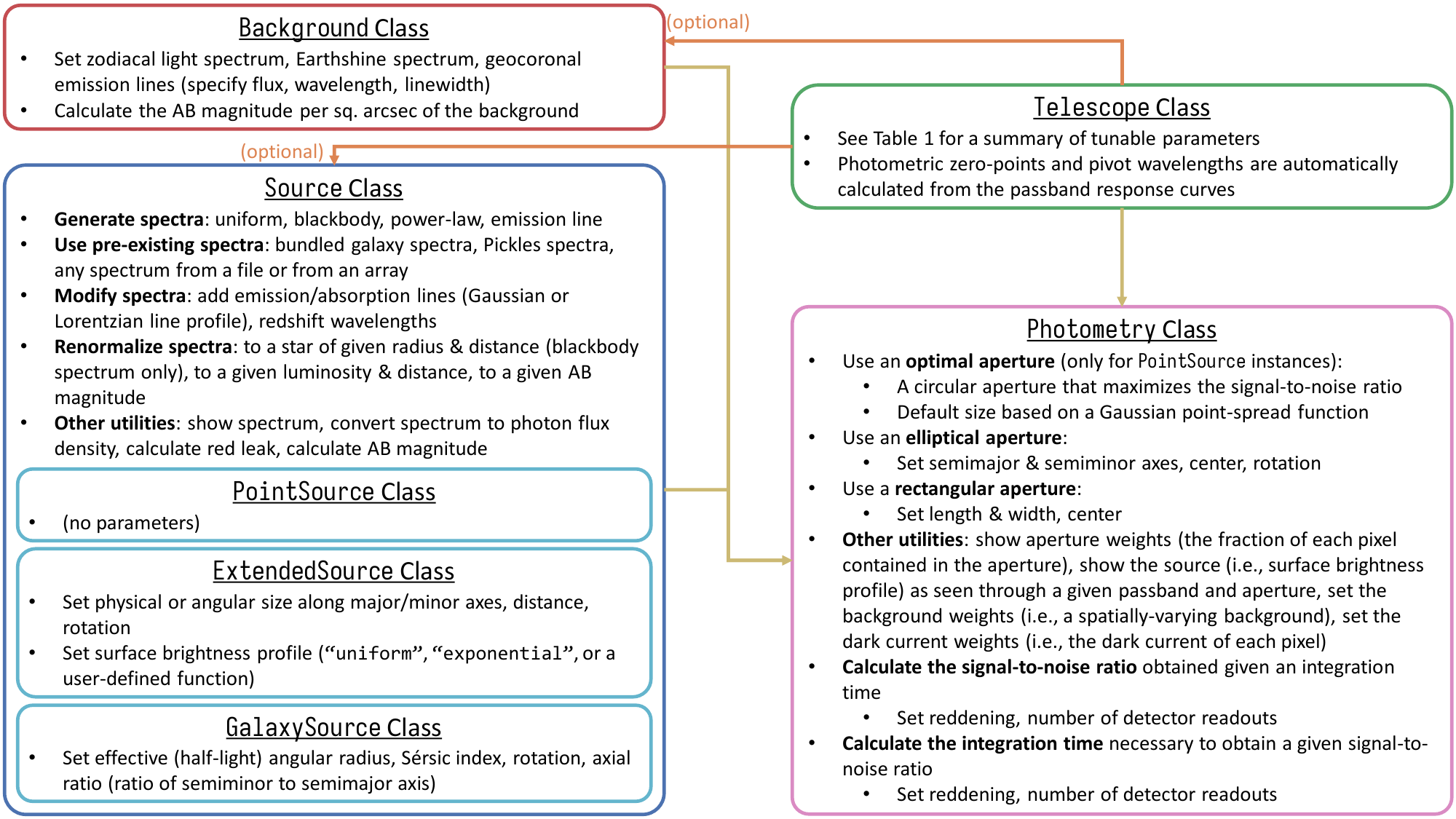}
  \caption{A flowchart describing the workflow associated with \cetc's photometry
   functionality. The most important parameters and methods of each class are included.}
  \label{fig:flowchart}
\end{figure*}

We summarize the workflow and the organization of the FORECASTOR photometry ETC in Figure
\ref{fig:flowchart}. At this moment, \cetc can only simulate single objects from start to
finish, however users can upload custom images (e.g., a crowded field) and use the ETC's
tools to obtain photometry and spectroscopy \citep{UVMOS} estimates.

\subsection{Graphical User Interface}\label{sec:gui}

While the full \cetc Python package is extremely flexible, creating \sn estimates from
\cetc requires some programming knowledge. Thus, we developed a graphical user interface
(GUI) for \cetc that requires no Python experience and can be accessed on any device with
an internet connection.

\begin{figure}[htb]
  \centering
  \includegraphics[width=\columnwidth]{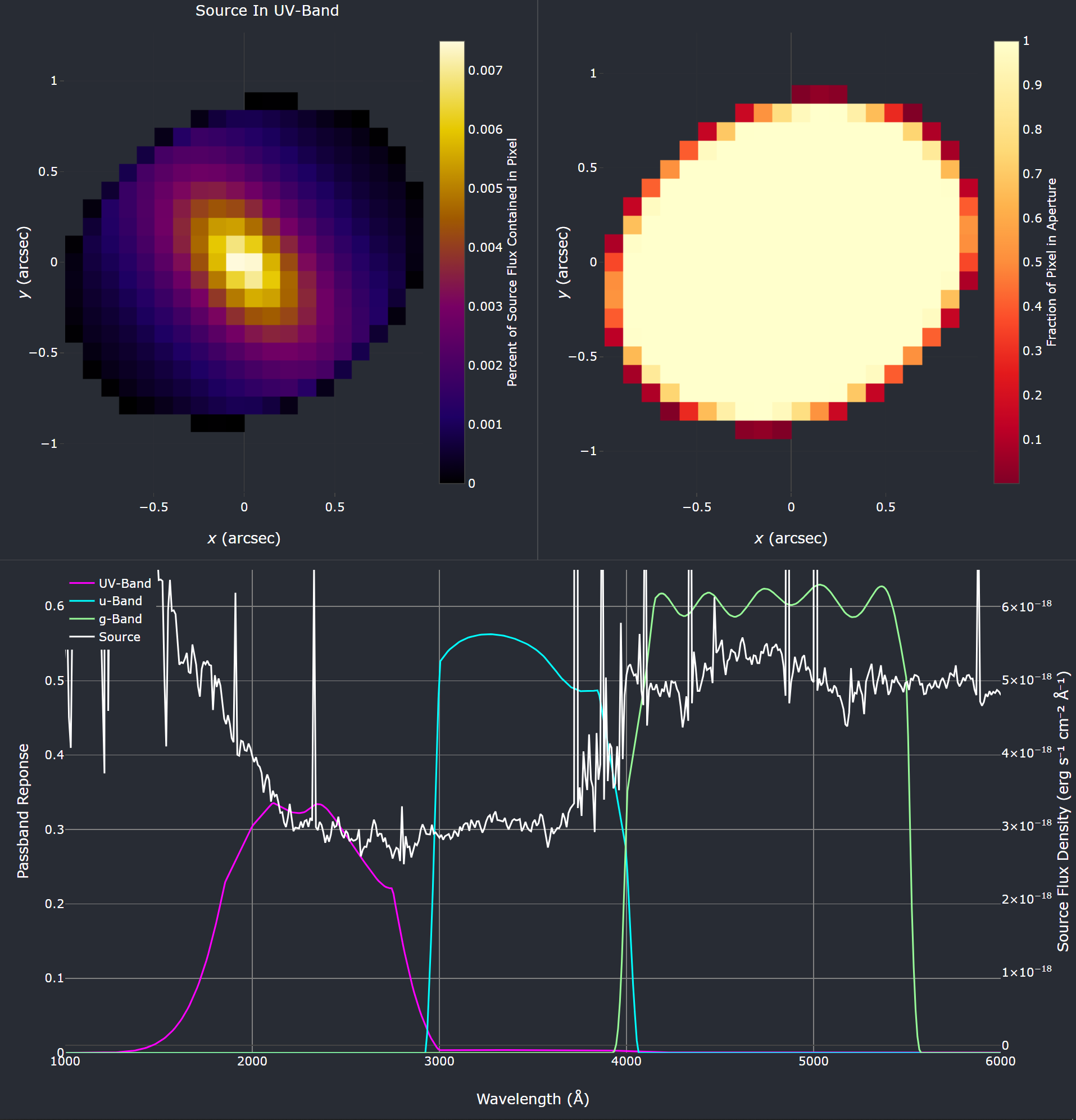}
  \caption{The right side of the graphical user interface, showing plots rendered by the
  FORECASTOR ETC for synthetic photometry calculations. These photometric calculations and
  mock aperture renderings update as the user changes and submits different parameters.
  The panels are resizable and all graphs are interactive (e.g., zoom and pan, show/hide
  different bands or spectra, obtain the value at a specific position in the graph,
  download graphs, etc.).}
  \label{fig:gui}
\end{figure}

The GUI is a web app developed in React that is currently hosted on the Canadian Advanced
Network for Astronomical Research (CANFAR\footnote{\url{https://www.canfar.net/en/}})
Science Portal, meaning it can even be used on a mobile device, such as a phone, with no
installation necessary. It is designed to mimic the procedure described previously, with
different tabs for each of the steps detailed in Section \ref{sec:etc}. Furthermore, the
interface and calculations update as the user changes and saves different observing
parameters. We show an example of this GUI in Figure \ref{fig:gui}.

\section{Example Science Cases}\label{sec:science_cases}

In order to demonstrate the functionality and utility of the FORECASTOR ETC, as well as
its results for the performance of the proposed {\it CASTOR} mission, we provide in the
following a few examples of preparatory calculations for planned {\it CASTOR} surveys.
Most examples here use a slightly more optimistic assumption of \(2.0\e\px^{-1}\) for the
read noise.

\subsection{Massive White Dwarfs}

The Magellanic Clouds provide an outstanding natural laboratory for studying populations
of stars, their fates, and their impact on their surrounding environment at a known
distance and subsolar (0.2--0.5\,Z$_{\odot}$) metallicities. There has been little
opportunity in the past, however, to study the Clouds systematically in the UV at high
spatial resolution. In particular, the resolution of previously-available wide-field
instruments has ranged from 1\farcs3 (\textit{AstroSat}/\textit{UVIT};
\citealt{tandon2020}) to 4\farcs5--5\farcs5 (\textit{GALEX}; \citealt{GALEX}). {\it
CASTOR}'s unique combination of sensitivity and spatial resolution will enable a new era
of resolved stellar population studies at UV and blue optical wavelengths.

A fundamental outstanding problem in stellar astrophysics remains identifying the maximum
initial mass of a star for which it will still leave behind a white dwarf (WD) at the end
of its nuclear-burning lifetime, and conversely, the minimum mass needed to undergo a core
collapse (Type II) supernova. The maximum mass of a WD is in itself well constrained at
$\simeq$1.37\Msun from theoretical considerations \citep{takahashi2013,althaus2022}, which
is consistent with the fact that none of the tens of thousands of spectroscopically
confirmed WDs have a mass exceeding $1.36\Msun$ \citep{kilic2021}. The maximum initial
stellar mass of a WD progenitor, thought to lie between 8 and $11\Msun$
\citep{weidemann1983,weidemann2000,siess2007,siess2010}, is much more uncertain. From a
theoretical perspective, it is hard to make much progress as this threshold is sensitive
to modelling assumptions concerning convection, overshoot, and mass loss during the late
phases of stellar evolution. Yet, determining this critical mass, and understanding how it
varies with metallicity, is essential to modelling the formation rates of compact objects
and gravitational wave events \citep{Giacobbo2019}, the nature of underluminous supernovae
\citep{Doherty2017}, and the chemical enrichment of the Universe \citep{Doherty2014}.

To better constrain this limit, recent studies have focused on finding massive WDs in
young clusters \citep{richer2021,miller2022}. In those populations, only massive stars
have had the time to evolve to the WD stage, and the cooling age of a WD can be compared
to the cluster age to establish the progenitor mass. Despite these efforts, few
constraints exist for progenitor masses in excess of $6\Msun$ and the WD initial-final
mass relation (IFMR) remains extremely uncertain in this high-mass regime
\citep{cumming2018}. A fundamental limitation of this approach is that there are very few
young clusters in the solar neighbourhood, and that searching further out in the Milky Way
is not promising due to confusion with field WDs. A new strategy recently demonstrated by
\cite{richer2022} consists of searching for massive WDs in young clusters in the
Magellanic Clouds, where galactic contaminants can be more easily excluded. With this
approach, \cite{richer2022} were able to identify five very hot ($T_{\rm eff} \simeq
150{,}000\K$) WD candidates in NGC 2164 using HST photometry.

\begin{figure}
    \centering
    \includegraphics[width=\columnwidth]{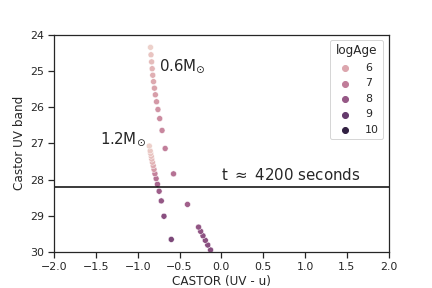}
    \caption{Evolution of a 0.6\Msun and a 1.2\Msun hydrogen-atmosphere WD in UV-u vs UV
    {\it CASTOR} magnitude Hertzsprung-Russell diagram. Points along each cooling track
    are coloured by age of the WD (as shown in the legend), while the horizontal line
    denotes the {\it CASTOR} limiting magnitude (\(\sn=5\)) in the UV-band given a 4200\s
    exposure. Here the UV magnitude includes \(E(B-V)=0.09\) for the LMC.}
    \label{fig:WDs}
\end{figure}

{\it CASTOR}'s unique UV sensitivity will allow a deeper and wider search for massive
cluster WDs in the Clouds. To quantify and confirm this, we make an estimate of the point
source sensitivity and compare with the WD cooling track in all bands assuming a distance
of 50\kpc to the Large Magellanic Cloud (LMC). Assuming for generality a flat AB magnitude
spectrum within each {\it CASTOR} band, Table \ref{SNRtable} provides the estimate from
the FORECASTOR ETC for the time needed to reach $\sn=5$ with a point source\footnote{A
version of this table as well as the code for generating it can be found
at:~\url{https://github.com/CASTOR-telescope/ETC_notebooks/blob/master/snr_table.ipynb}}.
In Figure \ref{fig:WDs}, we compare these limiting magnitudes to the evolution of a
hydrogen-atmosphere WD in the {\it CASTOR} ultraviolet HR diagram calculated using the
Montreal atmosphere models \citep{tremblay2011,bedard2020}.

\begin{deluxetable}{cDDD}[htbp]
  \tablecaption{Times needed to reach \(\sn=5\) for a given magnitude in a {\it CASTOR}
  band, assuming a flat continuum, \(E(B-V)=0.09\) for the LMC, and our default input
  telescope parameters.\label{SNRtable}}
  \tablehead{
  \colhead{\textit{CASTOR} Band} &
  \multicolumn{2}{r}{UV} &
  \multicolumn{2}{r}{u\hspace*{0.75ex}} &
  \multicolumn{2}{r}{g\hspace*{0.75ex}}
  \\[-6pt]
  \colhead{(AB mag)} &
  \multicolumn{2}{r}{(s)} &
  \multicolumn{2}{r}{(s)} &
  \multicolumn{2}{r}{(s)}
  }
  \decimals
  \startdata
  22.0 & 9.00    & 7.44    & 5.66 \\
  22.5 & 14.27   & 11.82   & 9.04 \\
  23.0 & 22.65   & 18.78   & 14.49 \\
  23.5 & 35.97   & 29.90   & 23.38 \\
  24.0 & 57.18   & 47.71   & 38.13 \\
  24.5 & 91.07   & 76.45   & 63.18 \\
  25.0 & 145.47  & 123.26  & 107.23 \\
  25.5 & 233.41  & 200.68  & 188.50 \\
  26.0 & 377.17  & 331.71  & 347.72 \\
  26.5 & 616.20  & 560.98  & 681.03 \\
  27.0 & 1023.90 & 981.08  & 1422.74 \\
  27.5 & 1745.22 & 1797.46 & 3151.41 \\
  27.6 & 1949.45 & 2042.54 & 3716.53 \\
  27.7 & 2181.01 & 2326.72 & 4390.34 \\
  28.0 & 3086.46 & 3492.19 & 7301.72 \\
  28.1 & 3478.63 & 4019.24 & 8673.27 \\
  28.2 & 3928.93 & 4637.94 & 10313.86 \\
  28.3 & 4447.37 & 5365.80 & 12277.18 \\
  28.5 & 5738.60 & 7236.97 & 17442.78 \\
  \enddata
\end{deluxetable}

The {\it CASTOR} Magellanic Clouds Survey will image the Clouds with 4200\,s exposures,
thereby reaching a depth of 28.2, 28.1, and 27.6 in the UV-, u- and g-bands, respectively.
This will enable the detection of the hot end of the WD cooling track in young Clouds
clusters (for UV $\sim$ 28.2, down to a WD $T_{\rm eff} \sim 40{,}000\,$K). The most
promising clusters to constrain the IFMR are those in the 40--80\,Myr age range, which is
when WDs are expected to have formed but for which we have not yet detected WDs in Milky
Way clusters. Promisingly, there are dozens of clusters within this age range in the
Clouds \citep{glatt2010}. Some may not be rich enough and others may have a crowded
background, but many of them are promising targets to identify young massive cluster WDs,
as demonstrated by \cite{richer2022}. Note that the detection of only a handful of WDs per
cluster is sufficient to provide useful constraints on the IFMR, since all that is
required is to positively identify the hot end of the WD cooling sequence.
Spectroscopically confirming the nature of WD candidates uncovered by this program will
remain out of reach for the foreseeable future, but WDs in the Clouds can be identified
with a very high degree of certainty solely based on their location in the {\it CASTOR} UV
HR diagram.

\subsection{M Star Flare Frequency Distributions}

All M stars exhibit long-lived phases of elevated magnetic activity and frequent flaring
throughout the first 0.5--5\,Gyr of their lifetimes \citep{Shkolnik_2014,Medina_2022}.
During their adolescence, M stars produce extreme levels of UV emission that drive
processes on orbiting planets, including atmospheric erosion \citep{Owen_2013},
photochemistry \citep{Hu_2014}, and impacts on surface habitability
\citep{Rugheimer_2015}. Characterizing the UV radiation environments of M stars both in
quiescence and during flares is critical for our understanding of exoplanetary atmospheric
processes and will be required to accurately interpret forthcoming biosignature detections
in exoplanetary atmospheres.

To date, \textit{GALEX} has served as the workhorse mission for characterizing M star UV
radiation environments. Its combination of wide sky coverage, long duration source
monitoring ($\geq 30$\,min), and capability to record time-tagged photon lists, allowed
\textit{GALEX} to characterize M stars' chromospheric UV emission
\citep{Shkolnik_2014,Schneider_2018} and UV flare rates \citep{Brasseur_2019,Jackman_2023}
as functions of stellar mass and age. \textit{CASTOR} will build upon the legacy of
\textit{GALEX} by leveraging its improved effective collecting area in the
NUV\footnote{\textit{CASTOR}'s expected effective collecting area is $\approx 50\times$
that of \textit{GALEX} in the NUV at 2200\AA \citep{CASTOR_SMS}.} and flexible observation
scheduling to conduct a deep time domain survey of M stars to measure their UV-u-g-band
flare frequency distributions (FFD; i.e., flare rate versus flare energy).

\textit{CASTOR}'s M star Legacy Survey will survey M star members of ten young moving
groups (YMG) in or near the telescope's continuous viewing zone (CVZ), as well as a
complement of field stars. In this way, the survey will sample the UV-u-g-band FFDs at
different stellar evolutionary stages from $\sim$2\,Myr to field ages, for both partially
and fully convective M stars. We used the FORECASTOR ETC to estimate the multi-band
photometric precision for each of our targets and to establish the depth of the M star
Legacy Survey. We selected targets by first cross-matching the \textit{Gaia} DR3 and 2MASS
catalogs to derive stellar masses based on the $M_{K_s}$-mass relation from
\citet{Mann_2019} and selecting stars with $M_\star< 0.65\Msun$. We then ran each target
through the BANYAN $\Sigma$ tool \citep{Gagne_2018} to determine YMG membership
probabilities based on \textit{Gaia} kinematics and assign stellar ages. We estimated each
star's UV-band AB magnitude using their J-band magnitudes and interpolating the
$F_{\mathrm{NUV}}/F_\mathrm{J}$ age sequences for partially convective early M stars
\citep[$>$0.35\Msun;][]{Shkolnik_2014} and fully convective mid-to-late M stars
\citep[$<$0.35\Msun;][]{Schneider_2018}. We then derive u- and g-band magnitudes by
scaling the semi-empirical HAZMAT spectral models, which accurately capture the
photospheric and chromospheric contributions to M stars' UV-optical SEDs as a function of
their mass and age \citep{Peacock_2020}.

\begin{figure}
    \centering
    \includegraphics[width=\hsize]{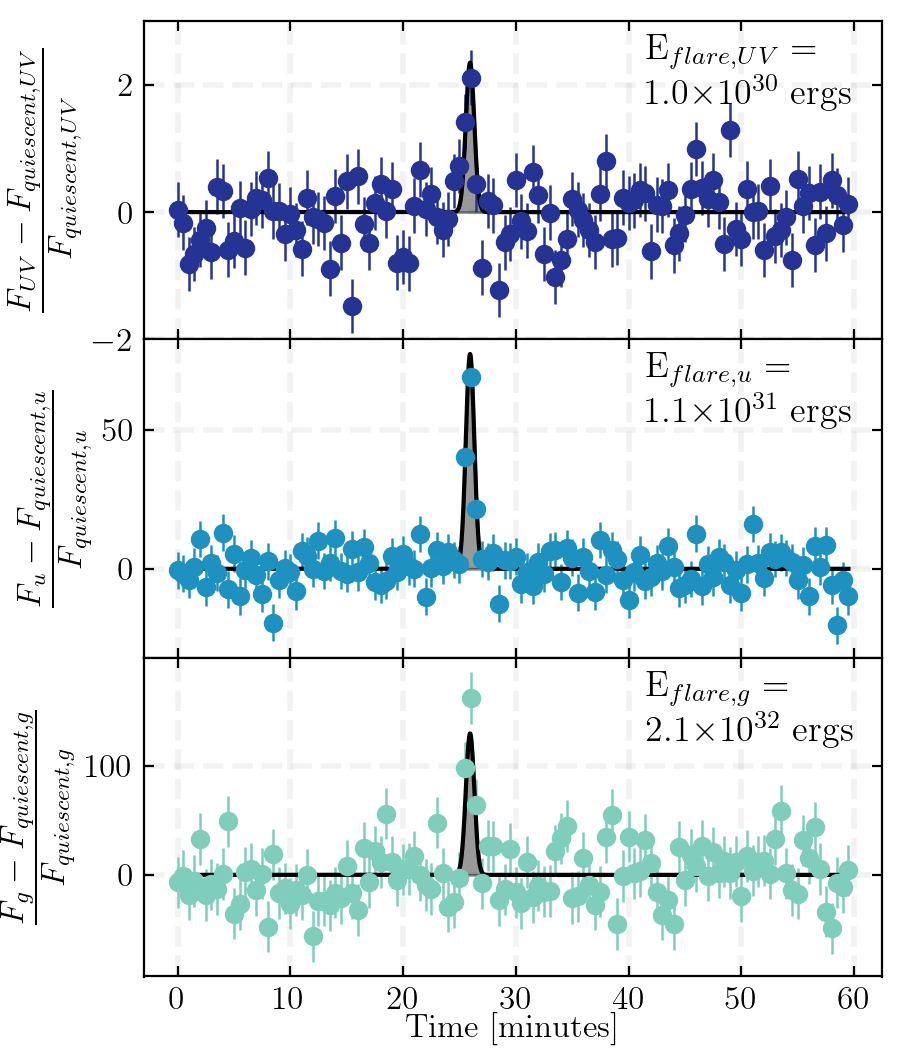}
    \caption{Synthetic {\it CASTOR} UV-u-g-band light curves for a randomly-selected flare
    star ($M_\star = 0.35\Msun$, Pleiades member at 112\,Myr; \citealt{Dahm_2015},
    $\{m_{\mathrm{AB,UV}}=25.8, m_{\mathrm{AB,u}}=23.2, m_{\mathrm{AB,g}}=20.0\}$). The
    cadence and photometric precision are based on the results from the FORECASTOR ETC for
    the fiducial observing strategy envisioned for CASTOR's M star Legacy Survey. Each
    light curve is offset and normalized by its quiescent flux level. The injected flare
    was sampled from the \textit{GALEX}/NUV FFD \citep{Jackman_2023} and is detected at
    $5.5\sigma$ in CASTOR's UV-band. The multi-band flare energies are calculated assuming
    a 9000\K blackbody and likely represent conservative estimates of the u- and g-band
    flare energies when compared to observations of M star flares with spectroscopic UV
    instruments such as \textit{HST}/\textit{COS} and \textit{HST}/\textit{STIS}
    \citep[e.g.][]{Loyd_2018,Kowalski_2019}.}
    \label{fig:flare}
\end{figure}

The results of the FORECASTOR ETC indicate that with a typical observing cadence of 21
seconds, {\it CASTOR} will achieve a \(\sn\!>\!10\) per UV-band exposure for more than
4000 M stars, which are roughly evenly split between partially versus fully convective
stars ($\lesssim$ 0.3\Msun) and between YMG members versus foreground field stars. This
level of photometric precision and observing cadence is sufficient to detect typical flare
energies in the UV-band \citep[i.e. $\sim$$10^{31}$\erg;][]{Jackman_2023} around stars
with $m_{\mathrm{AB,UV}}\lesssim 25.1$. The {\it CASTOR} M star Legacy Survey with this
limiting magnitude is about $40\times$ deeper than comparable M star surveys with
\textit{GALEX}/NUV and will have the capacity to monitor more than double the number of M
stars with observing durations sufficient to detect a statistically significant number of
flares \citep{Jackman_2023}. An example of multi-band light curves for a randomly-selected
Pleiades member, based on the FORECASTOR ETC predictions, is depicted in Figure
\ref{fig:flare}. Figure \ref{fig:flare} depicts typical light curves from the {\it CASTOR}
M star Legacy Survey and illustrates the detection of a typical flare above the star's
quiescent flux level.

\subsection{Measuring Proper Motions for Near-Field Cosmology}

We are entering a new era in the study of the Milky Way and nearby galaxies, and {\it
CASTOR} will play a vital role in characterizing known and newly discovered Galactic
substructures (stellar streams, globular clusters, dwarf galaxies---classical and
ultra-faint), in synergy with other wide-field observatories like \textit{Roman} and Vera
Rubin. In particular, thanks to its \textit{HST}-like spatial resolution, multi-epoch
imaging with {\it CASTOR} will enable precise measurements of proper motions for faint
stars, crucial for membership selection to study the stellar populations and density
distribution in these substructures but also for measuring systemic proper motions of
substructures and selecting targets for spectroscopic follow-up. Combined with its wide
field of view, the astrometric capabilities of {\it CASTOR} will open unique opportunities
in a range of subfields, but applications to near-field cosmology are perhaps the most
obvious given how {\it Gaia} has already revolutionized our understanding of the Milky Way
and its satellites, and how {\it CASTOR} can extend that exploration to fainter stars.
\textit{CASTOR}'s UV and blue-optical coverage would also provide better leverage on the
age and metallicity distribution of these faint stellar populations than the red-optical
and infrared observations of other missions like \textit{Roman} and \textit{Euclid}.

To predict the precision of proper motion measurements from \textit{CASTOR}, we assume
that the single-exposure astrometric precision for well-exposed point sources is 0.01\px
\citep[e.g.,][]{AndersonKing2006}, or about 1\,mas. This is consistent with current
experience on space-based platforms such as \textit{HST}, as long as a comparable level of
calibration activities are carried out. For each observation, this systematic error
($\sigma_{\rm sys}$) is added in quadrature with the random astrometric error
($\sigma_{\rm ast}$) such that the total astrometric error is given by $\sigma_{\rm tot} =
\sqrt{{\sigma_{\rm ast}}^2 + {\sigma_{\rm sys}}^2}$, where $\sigma_{\rm ast} = \sigma_{\rm
PSF}/{\rm (S/N)}$ and $\sigma_{\rm PSF}={\rm FWHM}/2.354$
\citep[e.g.,][]{Neuschaefer1995}, with FHWM taken to be 0\farcs15 for
\textit{CASTOR}\footnote{We assume that each observation is split in a sequence of four
dithered exposures to adequately sample the PSF given the detector's 0\farcs1 pixels.} and
the signal-to-noise ratio (S/N) of point sources calculated with the FORECASTOR ETC as
summarized below.

The precision of proper motion measurements ($\sigma_\mu$) depends on this total
astrometric error and improves when increasing the time baseline $T$ between the first and
last epoch of observations ($\sigma_\mu\!\propto\!1/T$). We assume here a typical time
baseline of four years between the first and last epoch given the mission lifetime of
\textit{CASTOR}, but note that further improvements in the precision of proper motions
would be achieved with epochs scheduled during an extended mission phase. It is also
assumed that at least one additional observation is obtained in-between the first and last
epoch to control systematics, based on experience with \textit{HST}. In principle, with
sufficient calibrations and platform stability, the astrometric precision of well-exposed
sources ($\sn\gtrsim200$) can be improved by repeated dithered observations as
$\sigma_{\rm tot} \propto 1/\sqrt{N}$, where $N$ is the number of dithered observations
per epoch \citep[e.g.,][]{WFIRST2019}. For most halo substructures, given their distances
of a few tens of kpc and the exposure time needed to reach this \sn, this strategy would
however be very time-consuming and not widely applicable. It may be worth exploring as
part of specific {\it CASTOR} programs, but for the proper motion error estimates
presented here we assume that this strategy is not used.

\begin{figure}
    \centering
    \includegraphics[width=\columnwidth]{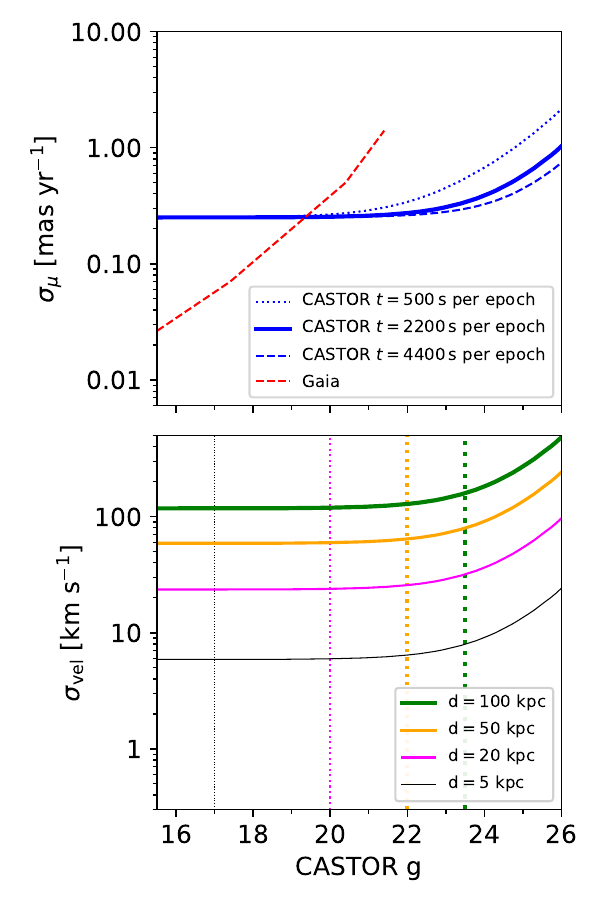}
    \caption{{\it Top panel:} Expected proper motion errors from {\it CASTOR} (in blue) as
    a function of {\it CASTOR} g-band magnitude for different total exposure times per
    epoch, assuming a maximum time baseline of four years. For comparison, typical proper
    motion uncertainties from \textit{Gaia} as a function of magnitude are also shown in
    red, where we converted from \textit{Gaia} G- to {\it CASTOR} g-band assuming the
    spectral energy distribution of a turnoff star for an old metal-poor stellar
    population. {\it Bottom panel:} expected plane-of-the-sky velocity precision from {\it
    CASTOR} as a function of {\it CASTOR} g-band magnitude for different distances (solid
    coloured lines), assuming an exposure time of 2200\s per epoch and a maximum time
    baseline of four years for proper motion measurements. The vertical dotted lines show
    the approximate magnitude of the main-sequence turnoff at those distances for an old
    metal-poor stellar population.}
    \label{fig:pm_precision}
\end{figure}

As the \sn is expected to be maximized in the g-band, our astrometric precision estimates
are based on the FORECASTOR ETC calculations in this band even though UV- and u-band
photometry would be obtained simultaneously (helping to characterize stellar populations).
We assume default FORECASTOR values for the sky background and an average
(\texttt{``avg''}) geocorononal emission flux. For each epoch, reaching $\sn=90$ at a
depth of ${\rm g}=23$ (about 3 magnitudes fainter than the {\it Gaia} limit) would limit
random astrometric errors and keep the total astrometric error for those faint stars close
to the lower limit set by the single-exposure systematic astrometric error of {\it CASTOR}
(see Figure \ref{fig:pm_precision}). This is achieved in a total exposure time of 2200\s
per epoch, which is what we assume for the reference survey described below.

As part of the {\it CASTOR Galactic Substructures Legacy Survey}, proper motions will be
measured for stars several magnitudes fainter than the {\it Gaia} limit (most Milky Way
halo stars are fainter than $\mathrm{G}=20$) in a large sample of targeted Milky Way
globular clusters, dwarf galaxies, and stellar streams. While these proper motions will in
general not be precise enough for studies of the internal kinematics of these systems (for
which $\lesssim1\km\s^{-1}$ or $\lesssim1\,\text{\textmu as}\,\mathrm{yr}^{-1}$ precision
would be required), they will be critical to detect and map very low surface brightness
features around these substructures by boosting sample sizes and statistical significance,
even at distances of several tens of kpc (where {\it CASTOR} can measure proper motions
with velocity uncertainties $<100\km\s^{-1}$ as shown in Figure \ref{fig:pm_precision},
which is generally sufficient for improved membership selection). This will provide a
unique view of their structure and composition, and it will be crucial to understand to
what extent their properties have been shaped by dark matter on sub-galactic scales and
tidal interaction with the Milky Way.

For example, {\it CASTOR} can probe the existence of small ``starless'' dark matter
sub-halos predicted by theory through their gravitational influence by looking for the
gaps they ``punch'' in thin stellar streams \citep[e.g.,][]{Erkal2016, Erkal2017,
Bovy2017, Price-Whelan2018, Bonaca2019}. For a sample of streams within a distance of
$\sim$20\kpc, {\it CASTOR} can measure velocities in proper motion with a precision of
$\sim$$30\km\s^{-1}$ or better for stars as faint as 3 magnitudes fainter than the {\it
Gaia} limit. At this depth, this proper motion precision while covering a large area of
the streams will be unprecedented. It will reveal a large number of new stream members and
be transformational for probing the morphology of the streams and perturbations from dark
matter sub-halos. This is important because current searches for the dynamical
perturbations of sub-halos on the morphology of streams are limited by knowledge of the
Milky Way background/foreground in the region of the streams and made difficult by
low-number statistics, leading to fluctuations in the star counts.

\subsection{Star Formation in Galaxies}\label{galaxies}

The UV emission from galaxies is usually dominated by massive stars, and is thus an
excellent tracer of cosmic star formation.  However, distant galaxies are faint in the UV
region and, beyond the local universe, subtend only a few square arcseconds on the sky.
\textit{GALEX} was the first mission with sufficient sensitivity and field of view to
allow a UV-based measurement of the star formation rate (SFR) density evolution. However,
due to its limited resolution ($\sim$5\arcsec) and sensitivity, \textit{GALEX} was only
able to detect fairly massive galaxies, and only out to redshifts of $z\!\sim\!1$
\citep[e.g.,][]{S+05}, even in its Deep Imaging Survey (covering $80\deg^2$ and reaching
\(\mathrm{NUV}=26\) for galaxies). Because galaxy formation depends sensitively on mass,
and peaks in activity around $z\!\sim\!2$ \citep{HB06,MD14}, a large fraction of the star
formation in the universe remains uncharted in the UV.

\textit{CASTOR} will provide critical information on both the recent star formation
histories of galaxies over cosmic time, and how this star formation is spatially
distributed within them. The wide area {\it CASTOR} surveys, such as the Deep and
Ultradeep surveys, will make it possible to put this within the context of the large-scale
structure of the Universe. The Deep survey will image 83\(\deg^2\) of the sky in six
contiguous regions overlapping with LSST/\emph{Euclid} Deep fields, reaching a 5\(\sigma\)
point source depth of \(m_\mathrm{AB}=29.75\) in the UV. The Ultradeep survey will extend
the sample to \(z>1.5\) and enable accurate pixel-by-pixel SFR estimates by having four
0.25\(\deg^2\) pointings with 10\(\times\) longer integrations. These will help
distinguish between the myriad feedback and dynamical mechanisms that can both stimulate
and hinder star formation.

We use the FORECASTOR ETC to estimate the depths and surface brightness limits of galaxy
photometry in the baseline Deep and Ultradeep surveys. The background includes an average
geocoronal contribution as described in Section \ref{sec:background}. In addition to the
standard {\it CASTOR} bands, we include an option for a broad-band filter that splits the
UV- and u-bands. An example of this, simply assuming a filter with sharp limits in the
centre of each band, is shown in Figure \ref{fig:filtercurves}.

\begin{figure}[b]
    \centering
    \includegraphics[width=\columnwidth]{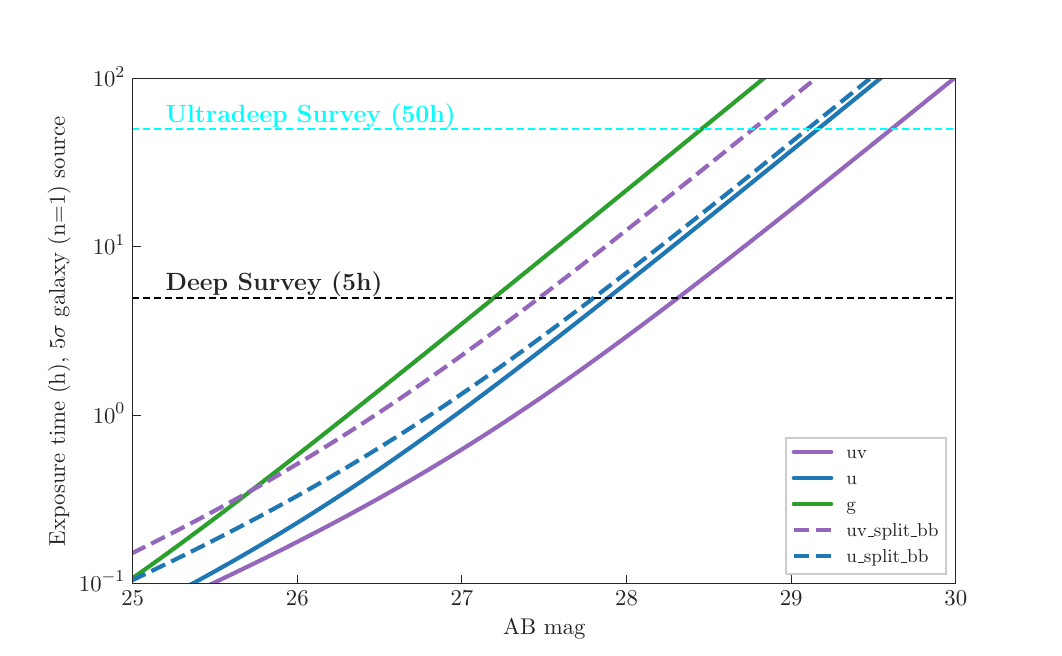}
    \caption{The exposure time required to reach $\sn\!=\!5$ as a function of AB
    magnitude, assuming a spiral galaxy at $z=1$ with $R_e=0\farcs25$ and $n=1$.}
    \label{fig:Gallimit}
\end{figure}

We assume a S\'ersic profile with an effective radius $R_e=0\farcs25$ and $n=1$,
appropriate for faint star-forming galaxies at $z=1$ \citep{vdW14}. The spectral energy
distribution is that of a local spiral galaxy, redshifted to $z=1$. For this calculation
the sources are assumed to be circular, and photometry is performed within a circular
aperture that is twice the effective radius. Figure \ref{fig:Gallimit} shows the time
required to reach $5\sigma$ as a function of AB magnitude in each of the five filters.
This is compared with the nominal depths of the Deep and Ultradeep surveys. Note that in
the actual surveys, the g-band will be observed twice:~once with the broadband filter in
place and once without. It will therefore have a longer exposure than in the other
filters.

\begin{figure}[h]
    \centering
    \includegraphics[width=\columnwidth]{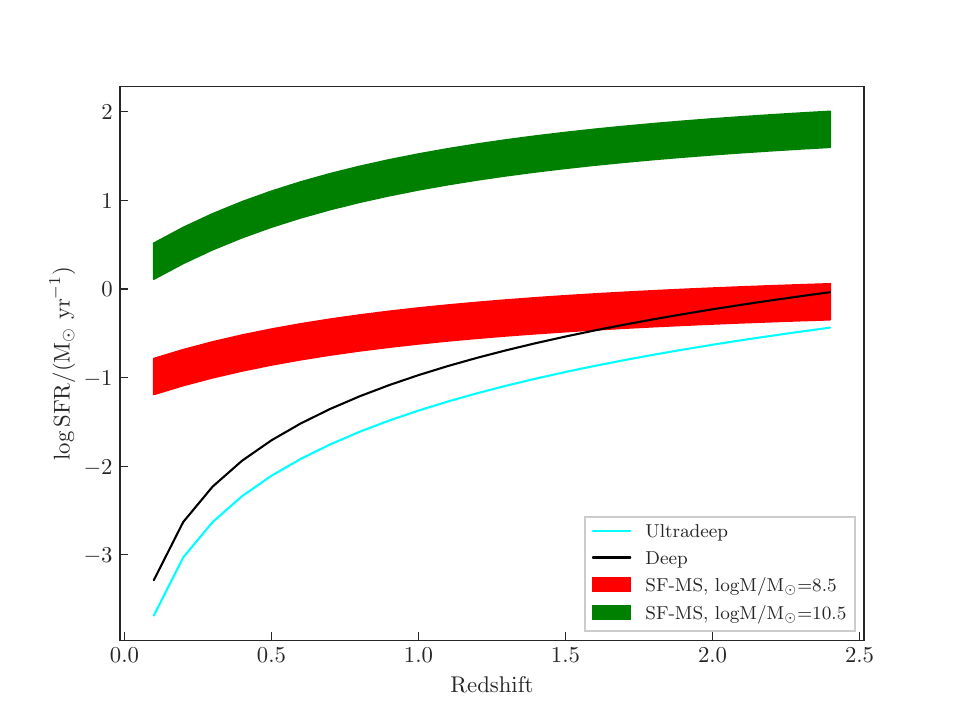}
    \caption{The locus of the star-forming main sequence and its 1$\sigma$ scatter, for
    galaxies with two stellar masses as indicated, is compared with the estimated
    5$\sigma$ extended source depths of the Deep (black line) and Ultradeep (cyan) curves,
    neglecting extinction. The main sequence parameterization is taken from \citet{SFRz}.}
    \label{fig:SFRz}
\end{figure}

To illustrate how these depths compare to the star formation rates of galaxies, we convert
UV magnitudes to SFR following the conversion of \citet{Kennicutt}:
\begin{equation}
\log{\mathrm{SFR}(\mathrm{M}_\odot\yr^{-1})}
=-27.85+\log{L_\nu (\mathrm{erg}\s^{-1}\Hz^{-1})},\label{eqn:sfr_from_L}
\end{equation}
where $L_\nu$ is the extinction-corrected luminosity at $150\lesssim
\lambda/\mathrm{nm}<280$. The SFR limits corresponding to the calculated UV depths of the
Deep and Ultradeep surveys are shown in Figure~\ref{fig:SFRz}.  These are compared with
the locus of the star-forming main sequence for galaxies with $\log(M/M_\odot)=8.5$ and
$\log(M/M_\odot)=10.5$, taken from \citet{SFRz}. Our estimated SFR limits neglect the
important effect of extinction and are thus optimistic. Nonetheless it is evident that
{\it CASTOR} will be able to detect almost all cosmic star formation out to cosmic noon
and, importantly, galaxies that lie well below the main sequence.

\begin{figure}
    \centering
    \includegraphics[width=\columnwidth]{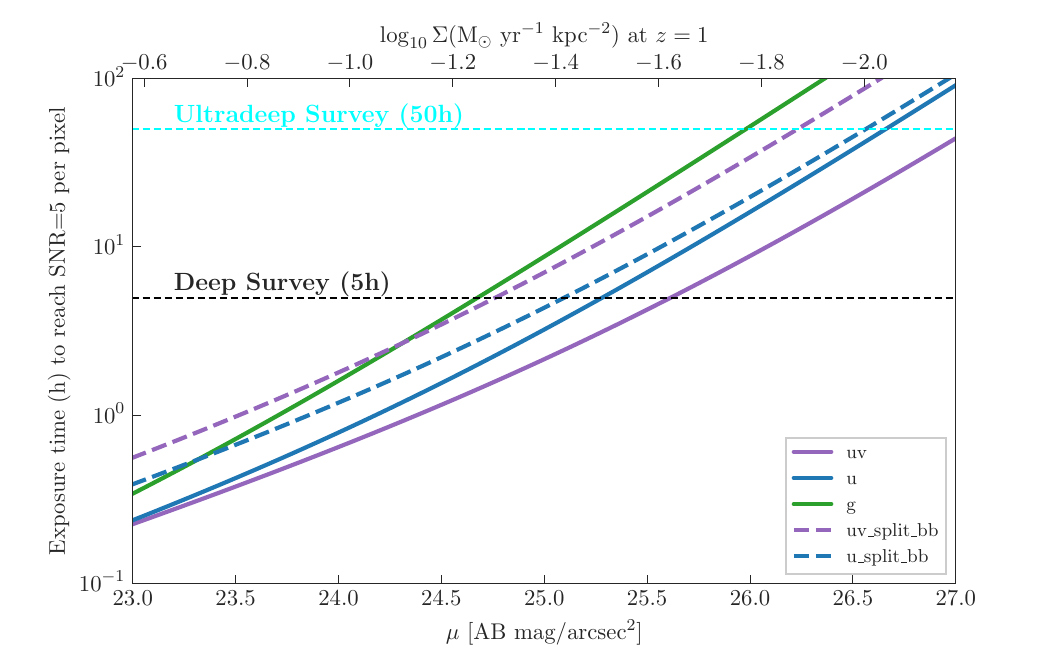}
    \caption{The time required to reach $\sn=5$ per pixel for a spiral galaxy at $z=1$
    with uniform surface brightness $\mu$, for each of the five {\it CASTOR} passbands.
    The top axis shows the corresponding physical star formation rate surface density,
    valid for $z=1$ and ignoring extinction, as described in the text.}
    \label{fig:SBlimit}
\end{figure}

The resolution and sensitivity of {\it CASTOR} enables an unprecedented opportunity to
study the spatial distribution of star formation in galaxies over wide fields, out to
$z=2$, using pixel-based spectral energy distribution (SED) fitting techniques
\citep[e.g.,][]{A+19,SS15,pixedfit}. To estimate the feasibility of this, we use the ETC
to find the exposure time required to reach $\sn=5$ per pixel at a given uniform surface
brightness level. This is shown in Figure \ref{fig:SBlimit} for each of the five filters,
relative to the proposed survey limits. In the UV, {\it CASTOR} will achieve this
sensitivity for a surface brightness of $\mu\approx25.6\,\mathrm{mag}\asec^{-2}$ in the
Deep survey, and $\mu\approx27.0\,\mathrm{mag}\asec^{-2}$ in the Ultradeep survey. For
$f_\nu$ measured in $\mathrm{erg}\s^{-1}\cm^{-2}\Hz^{-1}$ within an angular area
$\theta=1\asec^2$, we can convert to luminosity per area $A$ in $\mathrm{kpc}^2$ as
\begin{equation}
\begin{aligned}
\frac{L_\nu[\mathrm{erg}\s^{-1}\Hz^{-1}]}{A[\mathrm{kpc}^2]}
&=\frac{4\pi D_L^2f_\nu\left(3.086\times10^{21}\frac{\mathrm{cm}}{\mathrm{kpc}}\right)^2}{%
\theta[\mathrm{arcsec}^2]\left(\frac{D_A}{206265}\right)^2}\\
&=5.09\times 10^{54}\frac{f_\nu}{\theta}\left(\frac{D_L}{D_A}\right)^2\\
&=5.09\times 10^{54}\frac{f_\nu}{\theta}\left(1+z\right)^4
\end{aligned}
\end{equation}
where $D_L$ and $D_A$ are the luminosity and angular-diameter distances, respectively,
measured in units of kpc for a $\Lambda$CDM cosmology. Combining this with
Eq.~\eqref{eqn:sfr_from_L} yields a physical star formation rate surface density,
$\Sigma$:
\begin{equation}
\begin{aligned}
\log{\Sigma}&=-27.85+54.71-0.4\left(\mu+48.6\right)+4\log{\left(1+z\right)}\\
&=7.42-0.4\mu+4\log{\left(1+z\right),}
\end{aligned}
\end{equation}
for $\Sigma$ in units of $\mathrm{M}_\odot\yr^{-1}\kpc^{-2}$ and $\mu$ in AB
mag$\asec^{-2}$. The top axis of Figure \ref{fig:SBlimit} shows this corresponding SFR
density, for $z=1$.

With an ultradeep survey (50 hours per exposure) we will be able to constrain the SED in
the UV to \(27\,\mathrm{mag}\asec^{-2}\), which corresponds to the distant outskirts
($>$10\kpc) of typical galaxies \citep{Bouquin18}.

\begin{figure*}
  \centering
  \begin{minipage}[b]{0.49\textwidth}
    \centering
    \includegraphics[width=\textwidth]{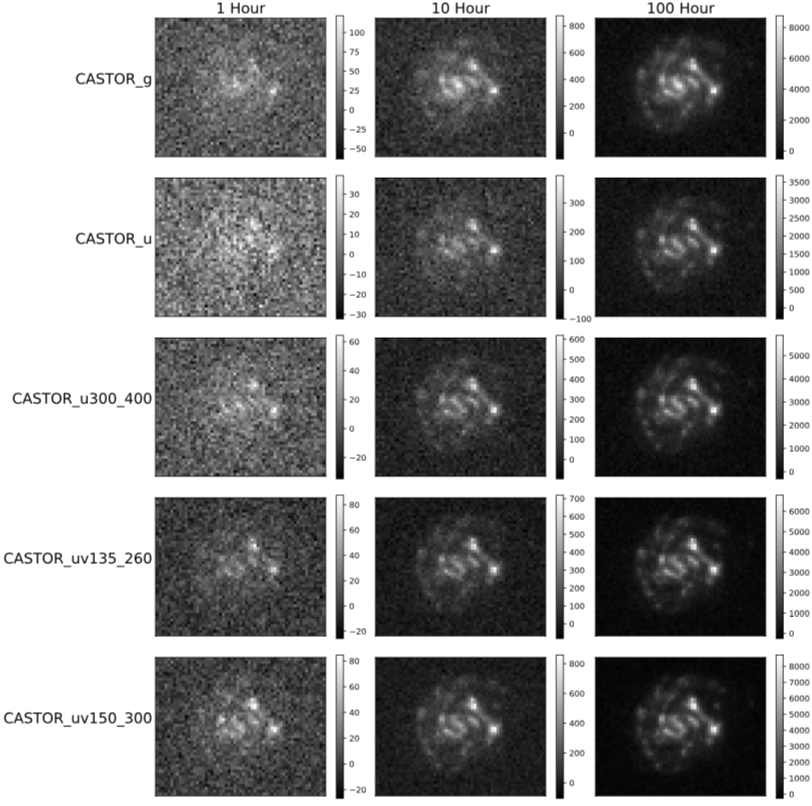}
    \vspace{0cm}

    (a)
  \end{minipage}
  \hfill
  \begin{minipage}[b]{0.49\textwidth}
    \centering
    \includegraphics[width=\textwidth]{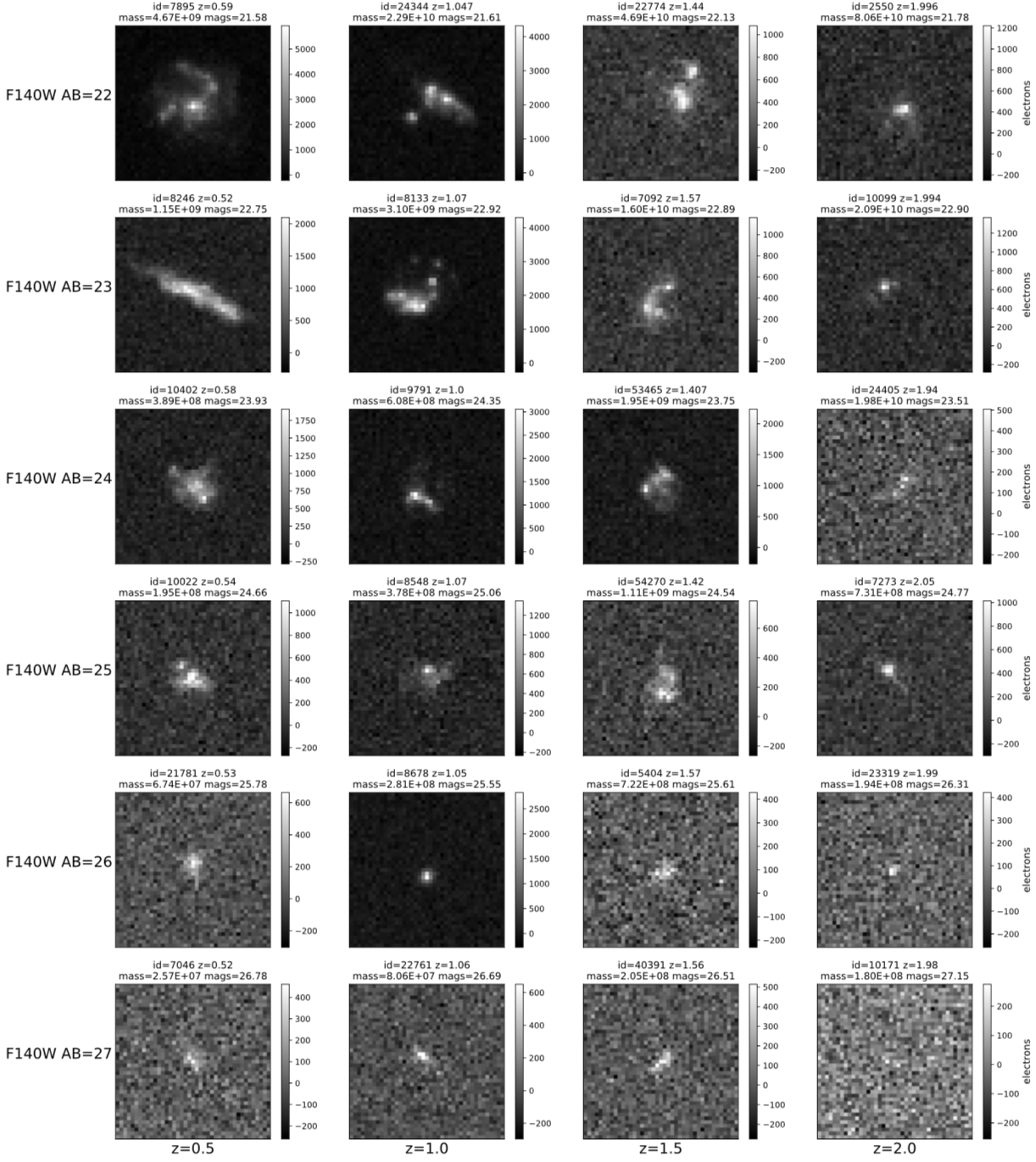}
    (b)
  \end{minipage}
  \caption{Left panels show the predicted \textit{CASTOR} images of a real \(z=0.6\)
   galaxy (taken from the Hubble UDF) at depths similar to those of the proposed
   \textit{CASTOR} Wide (left), Deep (middle) and Ultradeep (right) surveys. Right panels
   show representative UDF galaxies, chosen to span a grid of redshift and IR F140W
   magnitudes (a proxy for stellar mass) in simulated 100-hour \textit{CASTOR} UV
   observations.}
  \label{fig:castor_mock_imgs}
\end{figure*}

To illustrate what \textit{CASTOR} images of distant galaxies will look like, we show in
Figure \ref{fig:castor_mock_imgs} simulated g, u, and UV observations of galaxies in the
Hubble Ultra Deep Field (UDF). These images were produced by taking the spatially-resolved
spectral energy distribution (SED) fits of UDF galaxies from the work of \citet{Sorba2018}
and using them to model their pixel-by-pixel appearance in the \textit{CASTOR} filters.
These images were then degraded with noise values predicted by FORECASTOR. Panel (a) of
Figure \ref{fig:castor_mock_imgs} shows the result in UV, u, and g for a \(z=0.6\) UDF
galaxy under three illustrative exposure times (1, 10, and 100 hours). Panel (b) of Figure
\ref{fig:castor_mock_imgs} shows simulated 100-hour \textit{CASTOR} UV images of 24 real
UDF galaxies chosen to span a grid in redshift and brightness. In Figure
\ref{fig:castor_mock_imgs} panel (b), galaxies are shown as a function of IR magnitude
(\textit{Roman} F140W here, which can be regarded as a proxy for stellar mass) because for
extragalactic science, \textit{CASTOR} will leverage \textit{Roman} data.

\begin{figure}
  \centering
  \includegraphics[width=\columnwidth]{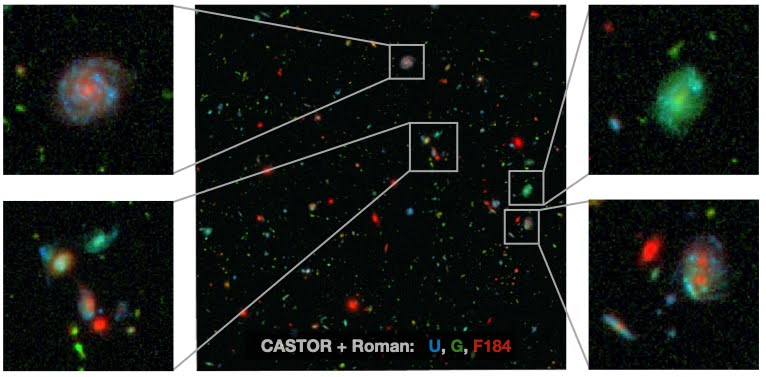}
  \caption{A simulation of a \(\sim\)1\farcm5\(\times\)1\farcm5 region within the Hubble
   Ultra Deep Field (UDF) in \textit{CASTOR} (u, g) and Roman (F184) filters. This shows
   \(\sim\)2.2\(\,\mathrm{arcmin}^2\), or just \(\sim\)0.06\% of the area that will be
   covered by the proposed \textit{CASTOR} Ultradeep Survey. The \textit{Roman} F184W
   image assumes an exposure time of 1 hour. The galaxy shown in the top left cutout is
   the same as that in the left panels of Figure \ref{fig:castor_mock_imgs}. While
   \textit{Roman}'s IR imaging (red) is sensitive to existing stellar mass,
   \textit{CASTOR} picks out regions of ongoing star formation. With comparable spatial
   resolution from the ultraviolet to the near-infrared, \textit{CASTOR} and
   \textit{Roman} working in concert will map out stellar populations and other physical
   parameters across galaxies out to the epoch of cosmic noon.}
  \label{fig:castor_roman_mock_udf}
\end{figure}

Figure \ref{fig:castor_roman_mock_udf} illustrates how \textit{CASTOR}'s sensitivity to
star formation in distant galaxies complements the sensitivity to existing stellar
populations that will be provided by space-borne IR observatories such as \textit{Roman}.
Here, the \textit{CASTOR} u- and g-band simulated images were produced by the procedure
described above and assumed 100-hour integrations, and the \textit{Roman} F184W image
followed a similar procedure but for a 1-hour integration.

\subsection{Time-Domain Studies of Active Galactic Nuclei}

Active galactic nuclei (AGN) are one of the most energetic systems in the
Universe---supermassive black holes (millions to billions of times more massive than our
Sun) surrounded by an ``accretion disk'' of ionized gas and dust---located at the centres
of massive galaxies. Accretion of matter onto the central black hole releases tremendous
energy over a broad range of the electromagnetic spectrum although AGN power peaks in the
UV. In addition, AGN are more variable in the UV than at optical or infrared wavelengths
\citep{macleod_etal_2010}. The variability timescales in the UV are also shorter than
those in the optical \citep{macleod_etal_2010}. For these reasons, the UV is a vital
regime for the study of AGN. \textit{CASTOR} will provide a unique window to the UV sky,
including the sensitivity required to perform UV observations of AGN, higher spatial
resolution to separate the central AGN from its host galaxy, and a slitless (grism) mode
that will allow taking spectra of a large number of AGN targets in a single field of view.

Time-domain studies of AGN take advantage of the variability in AGN to understand the
structure and kinematics of these systems.  AGN vary on several timescales
\citep{peterson_etal_1982}.  Tracking different timescales of variability allows us to
derive the sizes of the inner regions of AGN, that can then be used to estimate the
central black hole masses \citep[combining AGN sizes with the gas velocities given by
broad line widths from AGN spectra;][]{peterson_etal_2014}.  Determining black hole masses
of AGN over a wide range of redshifts is essential for understanding how supermassive
black holes grow over cosmic time.

\begin{figure}[ht]
    \centering
    \includegraphics[width=\columnwidth]{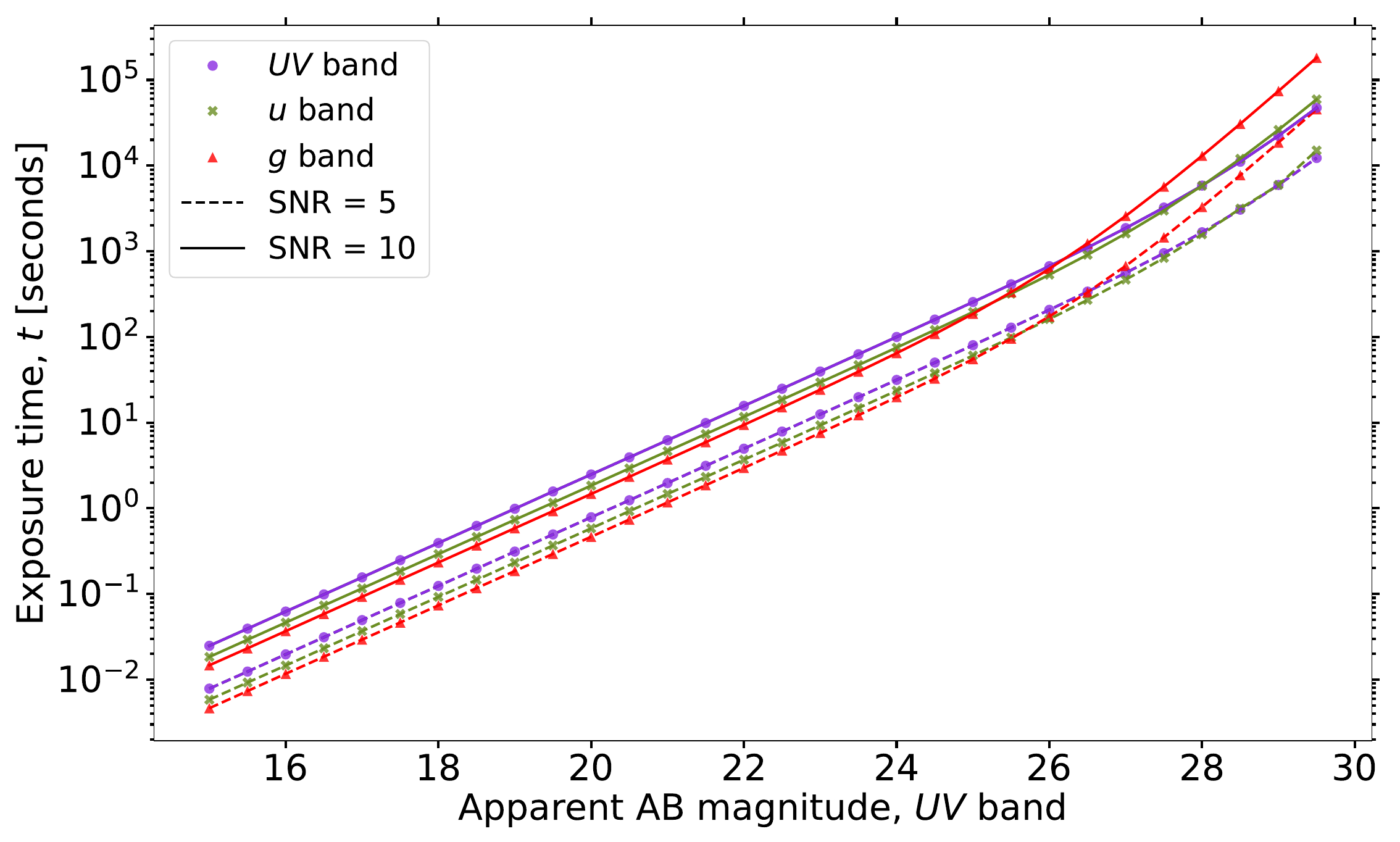}
    \caption{Exposure times required to reach \sn of 5 (dashed lines) and 10 (solid lines)
    in the UV- (purple circles), u- (green crosses), and g- (red triangles) bands for
    active galactic nuclei (AGN) observations.  The exposure times are calculated for a
    typical AGN spectrum in the rest frame normalized to AB magnitudes in the UV-band.
    Brighter targets (\(m_\mathrm{UV} \lesssim 24.5\)) require longer exposures in the
    bluer UV-band compared the redder g-band.}
    \label{fig:agn_exposure_times}
\end{figure}

Time-series analysis requires repeat observations of target AGN with sufficient \sn to
track their variability over a certain time period.  We used the FORECASTOR ETC to
calculate the exposure times required to reach \sn values of 5 and 10 for AGN with a range
of AB magnitudes in the UV-band of {\em CASTOR}, as shown in Figure
\ref{fig:agn_exposure_times}.  We assumed a background composed of Earthshine, zodiacal
light, and average emission from the \oii geocoronal line, together with a template AGN
spectrum from \citet{shang_etal_2011} in the rest frame.

Figure \ref{fig:agn_exposure_times} indicates that for brighter AGN, we need longer
exposure times in bluer bands (i.e., purple circles for the UV-band in Figure
\ref{fig:agn_exposure_times}) compared to redder bands (i.e., red triangles for the
g-band) to reach a desired \sn; however, fainter AGN require longer exposures in redder
bands.  Thus, \textit{CASTOR} will probe the fainter AGN population more efficiently in UV
bandpasses than in the optical g-band.

\begin{figure}[ht]
    \centering
    \includegraphics[width=\columnwidth]{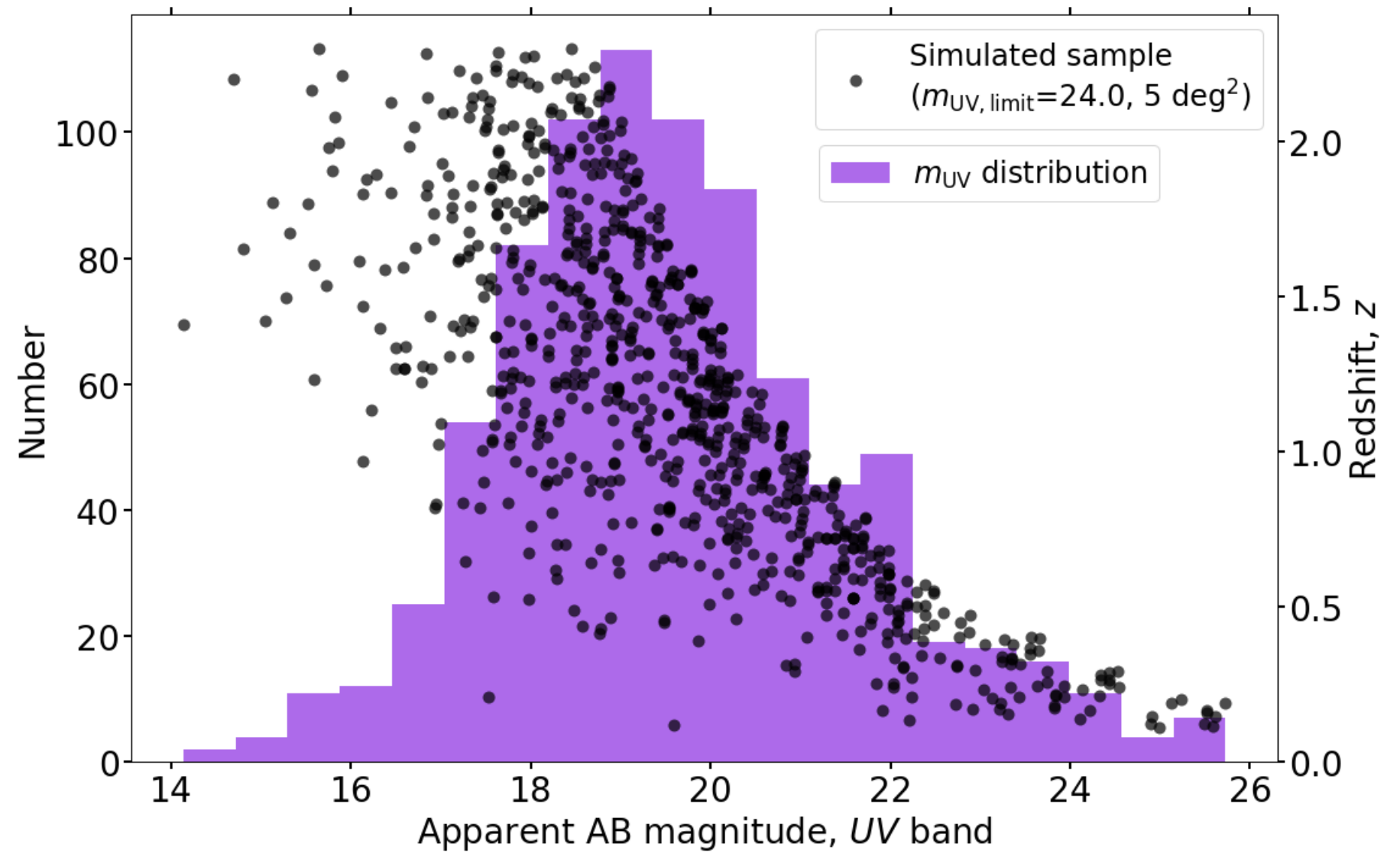}
    \caption{Apparent AB magnitude (\(m_\mathrm{UV}\)) distribution in the UV-band (purple
    histogram) illustrating number of AGN (left \(y\)-axis) that can be observed over a
    range of magnitudes (\(x\)-axis), over-plotted with individual AGN (black circles)
    over a range of redshifts (right \(y\)-axis) and \(m_\mathrm{UV}\) simulated in a sky
    area of 5\(\deg^2\) with a limiting \(m_\mathrm{UV}=24.0\).  In the simulated AGN
    sample, the objects at lower redshifts appear fainter, implying the need for several
    exposures to achieve desirable \sn values for those targets (see also discussion in
    the text).}
    \label{fig:agn_simulated_sample}
\end{figure}

With \textit{CASTOR}, we would need a range of different exposure times to probe AGN over
a wide redshift range.  Figure~\ref{fig:agn_simulated_sample} displays a realistic AGN
sample (black circles), obtained from the AGN UV luminosity function
\citep{kulkarni_etal_2019}, in the redshift-apparent AB magnitude space.  A fraction of
this sample extends to higher magnitudes ($m_\mathrm{UV}\geq21$; fainter targets).
While a \sn of 5 can be reached with shorter exposures for brighter targets, achieving
sufficient \sn to detect variability in fainter AGN would require considerably longer
exposures.  In such cases, stacking shorter exposures presents a way to build up to the
higher \sn required for fainter AGN observations.

With an AGN legacy survey, we aim to probe a unique luminosity-redshift parameter space
for AGN time-domain science that includes a significantly large number of objects at
low-to-medium luminosities in the redshift range of \(0.3 \lesssim z \lesssim 2.3\).

\section{Conclusion}\label{sec:concl}

The proposed {\it CASTOR} mission will be transformative, unveiling the ultraviolet and
blue-optical sky at high resolution across an enormous field of view. The first of the
FORECASTOR tools described above allows users in the community and prospective
collaborators to evaluate how {\it CASTOR} photometry will be able to benefit their
science case of interest. The ETC also allows users to compare and contrast
\textit{CASTOR}'s performance with any arbitrary instrument within the same software
environment, and indeed to develop associated multi-mission science cases using other
instruments, by adopting a generalized framework wherein the user can define a telescope,
background, or source object to arbitrary specifications. We have also developed a
flexible web interface for ``quick-look'' calculations or those that are not immediately
comfortable with working in Python. All of these tools are open access and available
ready-to-use through the Canadian Astronomy Data Centre's CANFAR Science Portal.

Using the FORECASTOR photometry ETC, we have shown a number of illustrative science cases
which highlight the science which {\it CASTOR} will be able to do, from assessing the
habitability of other solar systems, to tracing the motions of stars far fainter than
reached by \textit{Gaia}, to charting the rise and fall of star formation across the
history of the Universe.  In a companion work, \cite{UVMOS} describe an early
implementation of the UV multi-object spectrograph ETC tool, while \citet{Marshall2024}
and Noirot et al. (in prep.) describe the wide field image simulation and the grism ETC
tools, respectively. As the {\it CASTOR} mission looks ahead to the next phase in its
development, this work will be invaluable in defining the next steps towards launch in
$\sim$2030.

\begin{acknowledgments}

This paper is dedicated to the memory of our friend and colleague Harvey Richer, who led
foundational contributions to stellar and galactic astrophysics, to the Canadian space
astronomy community, and to fostering the next generation of Canadian scientists. The
authors also gratefully acknowledge support for this work provided by the Canadian Space
Agency.

\end{acknowledgments}

\bibliography{citations.bib}

\begin{thebibliography}{}
\expandafter\ifx\csname natexlab\endcsname\relax\def\natexlab#1{#1}\fi
\providecommand{\url}[1]{\href{#1}{#1}}
\providecommand{\dodoi}[1]{doi:~\href{http://doi.org/#1}{\nolinkurl{#1}}}
\providecommand{\doeprint}[1]{\href{http://ascl.net/#1}{\nolinkurl{http://ascl.net/#1}}}
\providecommand{\doarXiv}[1]{\href{https://arxiv.org/abs/#1}{\nolinkurl{https://arxiv.org/abs/#1}}}

\bibitem[{{Abdurro'uf} {et~al.}(2022){Abdurro'uf}, {Lin}, {Hirashita}, {Morishita}, {Tacchella}, {Akiyama}, {Takeuchi}, \& {Wu}}]{pixedfit}
{Abdurro'uf}, {Lin}, Y.-T., {Hirashita}, H., {et~al.} 2022, \apj, 926, 81, \dodoi{10.3847/1538-4357/ac439a}

\bibitem[{{Abraham} {et~al.}(1999){Abraham}, {Ellis}, {Fabian}, {Tanvir}, \& {Glazebrook}}]{A+19}
{Abraham}, R.~G., {Ellis}, R.~S., {Fabian}, A.~C., {Tanvir}, N.~R., \& {Glazebrook}, K. 1999, \mnras, 303, 641, \dodoi{10.1046/j.1365-8711.1999.02059.x}

\bibitem[{{Althaus} {et~al.}(2022){Althaus}, {Camisassa}, {Torres}, {Battich}, {C{\'o}rsico}, {Rebassa-Mansergas}, \& {Raddi}}]{althaus2022}
{Althaus}, L.~G., {Camisassa}, M.~E., {Torres}, S., {et~al.} 2022, \aap, 668, A58, \dodoi{10.1051/0004-6361/202244604}

\bibitem[{{Anderson} \& {King}(2006)}]{AndersonKing2006}
{Anderson}, J., \& {King}, I.~R. 2006, {PSFs, Photometry, and Astronomy for the ACS/WFC}, Instrument Science Report ACS 2006-01, 34 pages

\bibitem[{{B{\'e}dard} {et~al.}(2020){B{\'e}dard}, {Bergeron}, {Brassard}, \& {Fontaine}}]{bedard2020}
{B{\'e}dard}, A., {Bergeron}, P., {Brassard}, P., \& {Fontaine}, G. 2020, \apj, 901, 93, \dodoi{10.3847/1538-4357/abafbe}

\bibitem[{{Bessell} \& {Murphy}(2012)}]{Bessell2012}
{Bessell}, M., \& {Murphy}, S. 2012, \pasp, 124, 140, \dodoi{10.1086/664083}

\bibitem[{{Boggess} {et~al.}(1978){Boggess}, {Carr}, {Evans}, {Fischel}, {Freeman}, {Fuechsel}, {Klinglesmith}, {Krueger}, {Longanecker}, \& {Moore}}]{IUE}
{Boggess}, A., {Carr}, F.~A., {Evans}, D.~C., {et~al.} 1978, \nat, 275, 372, \dodoi{10.1038/275372a0}

\bibitem[{{Bonaca} {et~al.}(2019){Bonaca}, {Hogg}, {Price-Whelan}, \& {Conroy}}]{Bonaca2019}
{Bonaca}, A., {Hogg}, D.~W., {Price-Whelan}, A.~M., \& {Conroy}, C. 2019, \apj, 880, 38, \dodoi{10.3847/1538-4357/ab2873}

\bibitem[{{Bouquin} {et~al.}(2018){Bouquin}, {Gil de Paz}, {Mu{\~n}oz-Mateos}, {Boissier}, {Sheth}, {Zaritsky}, {Peletier}, {Knapen}, \& {Gallego}}]{Bouquin18}
{Bouquin}, A. Y.~K., {Gil de Paz}, A., {Mu{\~n}oz-Mateos}, J.~C., {et~al.} 2018, \apjs, 234, 18, \dodoi{10.3847/1538-4365/aaa384}

\bibitem[{{Bovy} {et~al.}(2017){Bovy}, {Erkal}, \& {Sanders}}]{Bovy2017}
{Bovy}, J., {Erkal}, D., \& {Sanders}, J.~L. 2017, \mnras, 466, 628, \dodoi{10.1093/mnras/stw3067}

\bibitem[{Bradley {et~al.}(2022)Bradley, Sipőcz, Robitaille, Tollerud, Vinícius, Deil, Barbary, Wilson, Busko, Donath, Günther, Cara, Lim, Meßlinger, Conseil, Bostroem, Droettboom, Bray, Bratholm, Barentsen, Craig, Rathi, Pascual, Perren, Georgiev, de~Val-Borro, Kerzendorf, Bach, Quint, \& Souchereau}]{photutils}
Bradley, L., Sipőcz, B., Robitaille, T., {et~al.} 2022, astropy/photutils:, 1.4.0,  Zenodo, \dodoi{10.5281/zenodo.6385735}

\bibitem[{{Branton} \& {Riley}(2021)}]{STISHandbook}
{Branton}, D., \& {Riley}, A. 2021, in STIS Instrument Handbook for Cycle 29 v.~20, Vol.~20, 20

\bibitem[{{Brasseur} {et~al.}(2019){Brasseur}, {Osten}, \& {Fleming}}]{Brasseur_2019}
{Brasseur}, C.~E., {Osten}, R.~A., \& {Fleming}, S.~W. 2019, \apj, 883, 88, \dodoi{10.3847/1538-4357/ab3df8}

\bibitem[{{Carruthers} \& {Page}(1972)}]{Apollo16FUV}
{Carruthers}, G.~R., \& {Page}, T. 1972, Science, 177, 788, \dodoi{10.1126/science.177.4051.788}

\bibitem[{{C{\^o}t{\'e}} {et~al.}(2019{\natexlab{a}}){C{\^o}t{\'e}}, {Roberto}, Michael, Peter, Maria, \& et~al.}]{CASTOR_SMS}
{C{\^o}t{\'e}}, P., {Roberto}, A., Michael, B., {et~al.} 2019{\natexlab{a}}, in CASTOR Science Maturation Study

\bibitem[{{C{\^o}t{\'e}} {et~al.}(2019{\natexlab{b}}){C{\^o}t{\'e}}, {Abraham}, {Balogh}, {Capak}, {Carlberg}, {Cowan}, {Djazovski}, {Drissen}, {Drout}, {Dupuis}, {Evans}, {Fantin}, {Ferrarese}, {Fraser}, {Gallagher}, {Girard}, {Gleisinger}, {Grandmont}, {Hall}, {Hellmich}, {Hardy}, {Harrison}, {Hlozek}, {Haggard}, {Henault-Brunet}, {Hutchings}, {Khatu}, {Kavelaars}, {Laurin}, {Lavigne}, {Lisman}, {Marois}, {McCabe}, {Metchev}, {Moutard}, {Netterfield}, {Nikzad}, {Ouellette}, {Pass}, {Parker}, {Pazder}, {Percival}, {Rhodes}, {Robert}, {Rowe}, {Sanchez-Janssen}, {Sivakoff}, {Shapiro}, {Sawicki}, {Scott}, {Van Waerbeke}, \& {Venn}}]{CASTOR}
{C{\^o}t{\'e}}, P., {Abraham}, B., {Balogh}, M., {et~al.} 2019{\natexlab{b}}, in Canadian Long Range Plan for Astronomy and Astrophysics White Papers, Vol. 2020, 18, \dodoi{10.5281/zenodo.3758463}

\bibitem[{{Cummings} {et~al.}(2018){Cummings}, {Kalirai}, {Tremblay}, {Ramirez-Ruiz}, \& {Choi}}]{cumming2018}
{Cummings}, J.~D., {Kalirai}, J.~S., {Tremblay}, P.~E., {Ramirez-Ruiz}, E., \& {Choi}, J. 2018, \apj, 866, 21, \dodoi{10.3847/1538-4357/aadfd6}

\bibitem[{{Dahm}(2015)}]{Dahm_2015}
{Dahm}, S.~E. 2015, \apj, 813, 108, \dodoi{10.1088/0004-637X/813/2/108}

\bibitem[{{de Jager} {et~al.}(1974){de Jager}, {Hoekstra}, {van der Hucht}, {Kamperman}, {Lamers}, {Hammerschlag}, {Werner}, \& {Emming}}]{TD1a}
{de Jager}, C., {Hoekstra}, R., {van der Hucht}, K.~A., {et~al.} 1974, \apss, 26, 207, \dodoi{10.1007/BF00642637}

\bibitem[{{Doherty} {et~al.}(2014){Doherty}, {Gil-Pons}, {Lau}, {Lattanzio}, \& {Siess}}]{Doherty2014}
{Doherty}, C.~L., {Gil-Pons}, P., {Lau}, H. H.~B., {Lattanzio}, J.~C., \& {Siess}, L. 2014, \mnras, 437, 195, \dodoi{10.1093/mnras/stt1877}

\bibitem[{{Doherty} {et~al.}(2017){Doherty}, {Gil-Pons}, {Siess}, \& {Lattanzio}}]{Doherty2017}
{Doherty}, C.~L., {Gil-Pons}, P., {Siess}, L., \& {Lattanzio}, J.~C. 2017, \pasa, 34, e056, \dodoi{10.1017/pasa.2017.52}

\bibitem[{{Dressel}(2021)}]{WFC3_handbook}
{Dressel}, L. 2021, in WFC3 Instrument Handbook for Cycle 29 v. 13, Vol.~13, 13

\bibitem[{{Erkal} {et~al.}(2016){Erkal}, {Belokurov}, {Bovy}, \& {Sanders}}]{Erkal2016}
{Erkal}, D., {Belokurov}, V., {Bovy}, J., \& {Sanders}, J.~L. 2016, \mnras, 463, 102, \dodoi{10.1093/mnras/stw1957}

\bibitem[{{Erkal} {et~al.}(2017){Erkal}, {Koposov}, \& {Belokurov}}]{Erkal2017}
{Erkal}, D., {Koposov}, S.~E., \& {Belokurov}, V. 2017, \mnras, 470, 60, \dodoi{10.1093/mnras/stx1208}

\bibitem[{{Fioc} \& {Rocca-Volmerange}(1997)}]{Fioc1997}
{Fioc}, M., \& {Rocca-Volmerange}, B. 1997, \aap, 326, 950.
\newblock \doarXiv{astro-ph/9707017}

\bibitem[{{Gagn{\'e}} {et~al.}(2018){Gagn{\'e}}, {Mamajek}, {Malo}, {Riedel}, {Rodriguez}, {Lafreni{\`e}re}, {Faherty}, {Roy-Loubier}, {Pueyo}, {Robin}, \& {Doyon}}]{Gagne_2018}
{Gagn{\'e}}, J., {Mamajek}, E.~E., {Malo}, L., {et~al.} 2018, \apj, 856, 23, \dodoi{10.3847/1538-4357/aaae09}

\bibitem[{{Giacobbo} \& {Mapelli}(2019)}]{Giacobbo2019}
{Giacobbo}, N., \& {Mapelli}, M. 2019, \mnras, 482, 2234, \dodoi{10.1093/mnras/sty2848}

\bibitem[{{Glatt} {et~al.}(2010){Glatt}, {Grebel}, \& {Koch}}]{glatt2010}
{Glatt}, K., {Grebel}, E.~K., \& {Koch}, A. 2010, \aap, 517, A50, \dodoi{10.1051/0004-6361/201014187}

\bibitem[{{Glover} {et~al.}(2022){Glover}, {Cheng}, {Woods}, {C{\^o}t{\'e}}, {Venn}, {Pazder}, {Hutchings}, \& {Blouin}}]{UVMOS}
{Glover}, J., {Cheng}, I., {Woods}, T.~E., {et~al.} 2022, in Society of Photo-Optical Instrumentation Engineers (SPIE) Conference Series, Vol. 12181, Society of Photo-Optical Instrumentation Engineers (SPIE) Conference Series, ed. J.-W.~A. {den Herder}, S.~{Nikzad}, \& K.~{Nakazawa}, 1218177, \dodoi{10.1117/12.2643009}

\bibitem[{{G{\'o}mez de Castro} {et~al.}(2021){G{\'o}mez de Castro}, {Barstow}, {Brosch}, {Cot{\'e}}, {France}, {Heap}, {Hutchings}, {Koriski}, {Murthy}, {Neiner}, {Roberge}, {Rom{\'a}n-Duval}, {Rowe}, {Sachkov}, {Schkolnik}, \& {Shustov}}]{GomezdeCastro}
{G{\'o}mez de Castro}, A.~I., {Barstow}, M.~A., {Brosch}, N., {et~al.} 2021, in Ultraviolet Astronomy and the Quest for the Origin of Life, ed. A.~I. {G{\'o}mez de Castro}, 115--160, \dodoi{10.1016/B978-0-12-819170-5.00004-X}

\bibitem[{{Green} {et~al.}(2012){Green}, {Froning}, {Osterman}, {Ebbets}, {Heap}, {Leitherer}, {Linsky}, {Savage}, {Sembach}, {Shull}, {Siegmund}, {Snow}, {Spencer}, {Stern}, {Stocke}, {Welsh}, {B{\'e}land}, {Burgh}, {Danforth}, {France}, {Keeney}, {McPhate}, {Penton}, {Andrews}, {Brownsberger}, {Morse}, \& {Wilkinson}}]{HSTCOS}
{Green}, J.~C., {Froning}, C.~S., {Osterman}, S., {et~al.} 2012, \apj, 744, 60, \dodoi{10.1088/0004-637X/744/1/60}

\bibitem[{{Gurzadyan} \& {Ohanesyan}(1972)}]{Orion}
{Gurzadyan}, G.~A., \& {Ohanesyan}, J.~B. 1972, \nat, 239, 90, \dodoi{10.1038/239090a0}

\bibitem[{{Hopkins} \& {Beacom}(2006)}]{HB06}
{Hopkins}, A.~M., \& {Beacom}, J.~F. 2006, \apj, 651, 142, \dodoi{10.1086/506610}

\bibitem[{{Hu} \& {Seager}(2014)}]{Hu_2014}
{Hu}, R., \& {Seager}, S. 2014, \apj, 784, 63, \dodoi{10.1088/0004-637X/784/1/63}

\bibitem[{{Jackman} {et~al.}(2023){Jackman}, {Shkolnik}, {Million}, {Fleming}, {Richey-Yowell}, \& {Loyd}}]{Jackman_2023}
{Jackman}, J. A.~G., {Shkolnik}, E.~L., {Million}, C., {et~al.} 2023, \mnras, 519, 3564, \dodoi{10.1093/mnras/stac3135}

\bibitem[{{Kennicutt}(1998)}]{Kennicutt}
{Kennicutt}, Robert~C., J. 1998, \araa, 36, 189, \dodoi{10.1146/annurev.astro.36.1.189}

\bibitem[{{Kilic} {et~al.}(2021){Kilic}, {Bergeron}, {Blouin}, \& {B{\'e}dard}}]{kilic2021}
{Kilic}, M., {Bergeron}, P., {Blouin}, S., \& {B{\'e}dard}, A. 2021, \mnras, 503, 5397, \dodoi{10.1093/mnras/stab767}

\bibitem[{{Kowalski} {et~al.}(2019){Kowalski}, {Wisniewski}, {Hawley}, {Osten}, {Brown}, {Fari{\~n}a}, {Valenti}, {Brown}, {Xilouris}, {Schmidt}, \& {Johns-Krull}}]{Kowalski_2019}
{Kowalski}, A.~F., {Wisniewski}, J.~P., {Hawley}, S.~L., {et~al.} 2019, \apj, 871, 167, \dodoi{10.3847/1538-4357/aaf058}

\bibitem[{{Kulkarni} {et~al.}(2019){Kulkarni}, {Worseck}, \& {Hennawi}}]{kulkarni_etal_2019}
{Kulkarni}, G., {Worseck}, G., \& {Hennawi}, J.~F. 2019, \mnras, 488, 1035, \dodoi{10.1093/mnras/stz1493}

\bibitem[{{Kumar} {et~al.}(2012){Kumar}, {Ghosh}, {Hutchings}, {Kamath}, {Kathiravan}, {Mahesh}, {Murthy}, {Nagbhushana}, {Pati}, {Rao}, {Rao}, {Sriram}, \& {Tandon}}]{UVIT}
{Kumar}, A., {Ghosh}, S.~K., {Hutchings}, J., {et~al.} 2012, in Society of Photo-Optical Instrumentation Engineers (SPIE) Conference Series, Vol. 8443, Space Telescopes and Instrumentation 2012: Ultraviolet to Gamma Ray, ed. T.~{Takahashi}, S.~S. {Murray}, \& J.-W.~A. {den Herder}, 84431N, \dodoi{10.1117/12.924507}

\bibitem[{{Linsky}(2018)}]{Linsky}
{Linsky}, J.~L. 2018, \apss, 363, 101, \dodoi{10.1007/s10509-018-3319-9}

\bibitem[{{Loyd} {et~al.}(2018){Loyd}, {Shkolnik}, {Schneider}, {Barman}, {Meadows}, {Pagano}, \& {Peacock}}]{Loyd_2018}
{Loyd}, R.~O.~P., {Shkolnik}, E.~L., {Schneider}, A.~C., {et~al.} 2018, \apj, 867, 70, \dodoi{10.3847/1538-4357/aae2ae}

\bibitem[{{MacLeod} {et~al.}(2010){MacLeod}, {Ivezi{\'c}}, {Kochanek}, {Koz{\l}owski}, {Kelly}, {Bullock}, {Kimball}, {Sesar}, {Westman}, {Brooks}, {Gibson}, {Becker}, \& {de Vries}}]{macleod_etal_2010}
{MacLeod}, C.~L., {Ivezi{\'c}}, {\v Z}., {Kochanek}, C.~S., {et~al.} 2010, \apj, 721, 1014, \dodoi{10.1088/0004-637X/721/2/1014}

\bibitem[{{Madau} \& {Dickinson}(2014)}]{MD14}
{Madau}, P., \& {Dickinson}, M. 2014, \araa, 52, 415, \dodoi{10.1146/annurev-astro-081811-125615}

\bibitem[{{Mann} {et~al.}(2019){Mann}, {Dupuy}, {Kraus}, {Gaidos}, {Ansdell}, {Ireland}, {Rizzuto}, {Hung}, {Dittmann}, {Factor}, {Feiden}, {Martinez}, {Ru{\'\i}z-Rodr{\'\i}guez}, \& {Thao}}]{Mann_2019}
{Mann}, A.~W., {Dupuy}, T., {Kraus}, A.~L., {et~al.} 2019, \apj, 871, 63, \dodoi{10.3847/1538-4357/aaf3bc}

\bibitem[{{Marshall} {et~al.}(2024){Marshall}, {Amen}, {Woods}, {Cote}, {Yung}, {Amenouche}, {Pass}, {Balogh}, {Salim}, \& {Moutard}}]{Marshall2024}
{Marshall}, M.~A., {Amen}, L., {Woods}, T.~E., {et~al.} 2024, arXiv e-prints, arXiv:2402.17163, \dodoi{10.48550/arXiv.2402.17163}

\bibitem[{{Martin} {et~al.}(2005){Martin}, {Fanson}, {Schiminovich}, {Morrissey}, {Friedman}, {Barlow}, {Conrow}, {Grange}, {Jelinsky}, {Milliard}, {Siegmund}, {Bianchi}, {Byun}, {Donas}, {Forster}, {Heckman}, {Lee}, {Madore}, {Malina}, {Neff}, {Rich}, {Small}, {Surber}, {Szalay}, {Welsh}, \& {Wyder}}]{GALEX}
{Martin}, D.~C., {Fanson}, J., {Schiminovich}, D., {et~al.} 2005, \apjl, 619, L1, \dodoi{10.1086/426387}

\bibitem[{{Mason} {et~al.}(2001){Mason}, {Breeveld}, {Much}, {Carter}, {Cordova}, {Cropper}, {Fordham}, {Huckle}, {Ho}, {Kawakami}, {Kennea}, {Kennedy}, {Mittaz}, {Pandel}, {Priedhorsky}, {Sasseen}, {Shirey}, {Smith}, \& {Vreux}}]{XMM_OM}
{Mason}, K.~O., {Breeveld}, A., {Much}, R., {et~al.} 2001, \aap, 365, L36, \dodoi{10.1051/0004-6361:20000044}

\bibitem[{{Medina} {et~al.}(2022){Medina}, {Winters}, {Irwin}, \& {Charbonneau}}]{Medina_2022}
{Medina}, A.~A., {Winters}, J.~G., {Irwin}, J.~M., \& {Charbonneau}, D. 2022, \apj, 935, 104, \dodoi{10.3847/1538-4357/ac77f9}

\bibitem[{{Mighell}(1999)}]{Mighell1999}
{Mighell}, K.~J. 1999, in Astronomical Society of the Pacific Conference Series, Vol. 189, Precision CCD Photometry, ed. E.~R. {Craine}, D.~L. {Crawford}, \& R.~A. {Tucker}, 50

\bibitem[{{Miller} {et~al.}(2022){Miller}, {Caiazzo}, {Heyl}, {Richer}, \& {Tremblay}}]{miller2022}
{Miller}, D.~R., {Caiazzo}, I., {Heyl}, J., {Richer}, H.~B., \& {Tremblay}, P.-E. 2022, \apjl, 926, L24, \dodoi{10.3847/2041-8213/ac50a5}

\bibitem[{{Moos} {et~al.}(2000){Moos}, {Cash}, {Cowie}, {Davidsen}, {Dupree}, {Feldman}, {Friedman}, {Green}, {Green}, {Gry}, {Hutchings}, {Jenkins}, {Linsky}, {Malina}, {Michalitsianos}, {Savage}, {Shull}, {Siegmund}, {Snow}, {Sonneborn}, {Vidal-Madjar}, {Willis}, {Woodgate}, {York}, {Ake}, {Andersson}, {Andrews}, {Barkhouser}, {Bianchi}, {Blair}, {Brownsberger}, {Cha}, {Chayer}, {Conard}, {Fullerton}, {Gaines}, {Grange}, {Gummin}, {Hebrard}, {Kriss}, {Kruk}, {Mark}, {McCarthy}, {Morbey}, {Murowinski}, {Murphy}, {Oegerle}, {Ohl}, {Oliveira}, {Osterman}, {Sahnow}, {Saisse}, {Sembach}, {Weaver}, {Welsh}, {Wilkinson}, \& {Zheng}}]{FUSE}
{Moos}, H.~W., {Cash}, W.~C., {Cowie}, L.~L., {et~al.} 2000, \apjl, 538, L1, \dodoi{10.1086/312795}

\bibitem[{{Neuschaefer} \& {Windhorst}(1995)}]{Neuschaefer1995}
{Neuschaefer}, L.~W., \& {Windhorst}, R.~A. 1995, \apjs, 96, 371, \dodoi{10.1086/192124}

\bibitem[{{Owen} \& {Wu}(2013)}]{Owen_2013}
{Owen}, J.~E., \& {Wu}, Y. 2013, \apj, 775, 105, \dodoi{10.1088/0004-637X/775/2/105}

\bibitem[{{Peacock} {et~al.}(2020){Peacock}, {Barman}, {Shkolnik}, {Loyd}, {Schneider}, {Pagano}, \& {Meadows}}]{Peacock_2020}
{Peacock}, S., {Barman}, T., {Shkolnik}, E.~L., {et~al.} 2020, \apj, 895, 5, \dodoi{10.3847/1538-4357/ab893a}

\bibitem[{{Peterson} {et~al.}(1982){Peterson}, {Foltz}, {Byard}, \& {Wagner}}]{peterson_etal_1982}
{Peterson}, B.~M., {Foltz}, C.~B., {Byard}, P.~L., \& {Wagner}, R.~M. 1982, \apjs, 49, 469, \dodoi{10.1086/190807}

\bibitem[{{Peterson} {et~al.}(2014){Peterson}, {Grier}, {Horne}, {Pogge}, {Bentz}, {De Rosa}, {Denney}, {Martini}, {Sergeev}, {Kaspi}, {Minezaki}, {Zu}, {Kochanek}, {Siverd}, {Shappee}, {Araya Salvo}, {Beatty}, {Bird}, {Bord}, {Borman}, {Che}, {Chen}, {Cohen}, {Dietrich}, {Doroshenko}, {Drake}, {Efimov}, {Free}, {Ginsburg}, {Henderson}, {King}, {Koshida}, {Mogren}, {Molina}, {Mosquera}, {Motohara}, {Nazarov}, {Okhmat}, {Pejcha}, {Rafter}, {Shields}, {Skowron}, {Skowron}, {Valluri}, {van Saders}, \& {Yoshii}}]{peterson_etal_2014}
{Peterson}, B.~M., {Grier}, C.~J., {Horne}, K., {et~al.} 2014, \apj, 795, 149, \dodoi{10.1088/0004-637X/795/2/149}

\bibitem[{{Pickles}(1998)}]{Pickles1998}
{Pickles}, A.~J. 1998, \pasp, 110, 863, \dodoi{10.1086/316197}

\bibitem[{{Popesso} {et~al.}(2023){Popesso}, {Concas}, {Cresci}, {Belli}, {Rodighiero}, {Inami}, {Dickinson}, {Ilbert}, {Pannella}, \& {Elbaz}}]{SFRz}
{Popesso}, P., {Concas}, A., {Cresci}, G., {et~al.} 2023, \mnras, 519, 1526, \dodoi{10.1093/mnras/stac3214}

\bibitem[{{Price-Whelan} \& {Bonaca}(2018)}]{Price-Whelan2018}
{Price-Whelan}, A.~M., \& {Bonaca}, A. 2018, \apjl, 863, L20, \dodoi{10.3847/2041-8213/aad7b5}

\bibitem[{{Pritchet} \& {Kline}(1981)}]{Pritchet1981}
{Pritchet}, C., \& {Kline}, M.~I. 1981, \aj, 86, 1859, \dodoi{10.1086/113065}

\bibitem[{{Richer} {et~al.}(2021){Richer}, {Caiazzo}, {Du}, {Grondin}, {Hegarty}, {Heyl}, {Kerr}, {Miller}, \& {Thiele}}]{richer2021}
{Richer}, H.~B., {Caiazzo}, I., {Du}, H., {et~al.} 2021, \apj, 912, 165, \dodoi{10.3847/1538-4357/abdeb7}

\bibitem[{{Richer} {et~al.}(2022){Richer}, {Cohen}, {Heyl}, {Kalirai}, {Caiazzo}, {Correnti}, {Cummings}, {Goudfrooij}, {Hansen}, {Peeples}, {Sabbi}, {Tremblay}, \& {Williams}}]{richer2022}
{Richer}, H.~B., {Cohen}, R.~E., {Heyl}, J., {et~al.} 2022, \apjl, 931, L20, \dodoi{10.3847/2041-8213/ac6585}

\bibitem[{{Roming} {et~al.}(2005){Roming}, {Kennedy}, {Mason}, {Nousek}, {Ahr}, {Bingham}, {Broos}, {Carter}, {Hancock}, {Huckle}, {Hunsberger}, {Kawakami}, {Killough}, {Koch}, {McLelland}, {Smith}, {Smith}, {Soto}, {Boyd}, {Breeveld}, {Holland}, {Ivanushkina}, {Pryzby}, {Still}, \& {Stock}}]{Swift_UVOT}
{Roming}, P. W.~A., {Kennedy}, T.~E., {Mason}, K.~O., {et~al.} 2005, \ssr, 120, 95, \dodoi{10.1007/s11214-005-5095-4}

\bibitem[{{Rugheimer} {et~al.}(2015){Rugheimer}, {Kaltenegger}, {Segura}, {Linsky}, \& {Mohanty}}]{Rugheimer_2015}
{Rugheimer}, S., {Kaltenegger}, L., {Segura}, A., {Linsky}, J., \& {Mohanty}, S. 2015, \apj, 809, 57, \dodoi{10.1088/0004-637X/809/1/57}

\bibitem[{{Schiminovich} {et~al.}(2005){Schiminovich}, {Ilbert}, {Arnouts}, {Milliard}, {Tresse}, {Le F{\`e}vre}, {Treyer}, {Wyder}, {Budav{\'a}ri}, {Zucca}, {Zamorani}, {Martin}, {Adami}, {Arnaboldi}, {Bardelli}, {Barlow}, {Bianchi}, {Bolzonella}, {Bottini}, {Byun}, {Cappi}, {Contini}, {Charlot}, {Donas}, {Forster}, {Foucaud}, {Franzetti}, {Friedman}, {Garilli}, {Gavignaud}, {Guzzo}, {Heckman}, {Hoopes}, {Iovino}, {Jelinsky}, {Le Brun}, {Lee}, {Maccagni}, {Madore}, {Malina}, {Marano}, {Marinoni}, {McCracken}, {Mazure}, {Meneux}, {Morrissey}, {Neff}, {Paltani}, {Pell{\`o}}, {Picat}, {Pollo}, {Pozzetti}, {Radovich}, {Rich}, {Scaramella}, {Scodeggio}, {Seibert}, {Siegmund}, {Small}, {Szalay}, {Vettolani}, {Welsh}, {Xu}, \& {Zanichelli}}]{S+05}
{Schiminovich}, D., {Ilbert}, O., {Arnouts}, S., {et~al.} 2005, \apjl, 619, L47, \dodoi{10.1086/427077}

\bibitem[{{Schneider} \& {Shkolnik}(2018)}]{Schneider_2018}
{Schneider}, A.~C., \& {Shkolnik}, E.~L. 2018, \aj, 155, 122, \dodoi{10.3847/1538-3881/aaaa24}

\bibitem[{{Shang} {et~al.}(2011){Shang}, {Brotherton}, {Wills}, {Wills}, {Cales}, {Dale}, {Green}, {Runnoe}, {Nemmen}, {Gallagher}, {Ganguly}, {Hines}, {Kelly}, {Kriss}, {Li}, {Tang}, \& {Xie}}]{shang_etal_2011}
{Shang}, Z., {Brotherton}, M.~S., {Wills}, B.~J., {et~al.} 2011, \apjs, 196, 2, \dodoi{10.1088/0067-0049/196/1/2}

\bibitem[{{Shkolnik} \& {Barman}(2014)}]{Shkolnik_2014}
{Shkolnik}, E.~L., \& {Barman}, T.~S. 2014, \aj, 148, 64, \dodoi{10.1088/0004-6256/148/4/64}

\bibitem[{{Siess}(2007)}]{siess2007}
{Siess}, L. 2007, \aap, 476, 893, \dodoi{10.1051/0004-6361:20078132}

\bibitem[{{Siess}(2010)}]{siess2010}
---. 2010, \aap, 512, A10, \dodoi{10.1051/0004-6361/200913556}

\bibitem[{{Sorba} \& {Sawicki}(2015)}]{SS15}
{Sorba}, R., \& {Sawicki}, M. 2015, \mnras, 452, 235, \dodoi{10.1093/mnras/stv1235}

\bibitem[{{Sorba} \& {Sawicki}(2018)}]{Sorba2018}
---. 2018, \mnras, 476, 1532, \dodoi{10.1093/mnras/sty186}

\bibitem[{{STScI Development Team}(2013)}]{pysynphot}
{STScI Development Team}. 2013, {pysynphot: Synthetic photometry software package}, Astrophysics Source Code Library, record ascl:1303.023.
\newblock \doeprint{1303.023}

\bibitem[{{Takahashi} {et~al.}(2013){Takahashi}, {Yoshida}, \& {Umeda}}]{takahashi2013}
{Takahashi}, K., {Yoshida}, T., \& {Umeda}, H. 2013, \apj, 771, 28, \dodoi{10.1088/0004-637X/771/1/28}

\bibitem[{{Tandon} {et~al.}(2020){Tandon}, {Postma}, {Joseph}, {Devaraj}, {Subramaniam}, {Barve}, {George}, {Ghosh}, {Girish}, {Hutchings}, {Kamath}, {Kathiravan}, {Kumar}, {Lancelot}, {Leahy}, {Mahesh}, {Mohan}, {Nagabhushana}, {Pati}, {Rao}, {Sankarasubramanian}, {Sriram}, \& {Stalin}}]{tandon2020}
{Tandon}, S.~N., {Postma}, J., {Joseph}, P., {et~al.} 2020, \aj, 159, 158, \dodoi{10.3847/1538-3881/ab72a3}

\bibitem[{{The LUVOIR Team}(2019)}]{LUVOIR}
{The LUVOIR Team}. 2019, arXiv e-prints, arXiv:1912.06219.
\newblock \doarXiv{1912.06219}

\bibitem[{{Tokunaga} \& {Vacca}(2005)}]{TV05}
{Tokunaga}, A.~T., \& {Vacca}, W.~D. 2005, \pasp, 117, 421, \dodoi{10.1086/429382}

\bibitem[{{Tremblay} {et~al.}(2011){Tremblay}, {Bergeron}, \& {Gianninas}}]{tremblay2011}
{Tremblay}, P.~E., {Bergeron}, P., \& {Gianninas}, A. 2011, \apj, 730, 128, \dodoi{10.1088/0004-637X/730/2/128}

\bibitem[{{van der Wel} {et~al.}(2014){van der Wel}, {Franx}, {van Dokkum}, {Skelton}, {Momcheva}, {Whitaker}, {Brammer}, {Bell}, {Rix}, {Wuyts}, {Ferguson}, {Holden}, {Barro}, {Koekemoer}, {Chang}, {McGrath}, {H{\"a}ussler}, {Dekel}, {Behroozi}, {Fumagalli}, {Leja}, {Lundgren}, {Maseda}, {Nelson}, {Wake}, {Patel}, {Labb{\'e}}, {Faber}, {Grogin}, \& {Kocevski}}]{vdW14}
{van der Wel}, A., {Franx}, M., {van Dokkum}, P.~G., {et~al.} 2014, \apj, 788, 28, \dodoi{10.1088/0004-637X/788/1/28}

\bibitem[{{Weidemann}(2000)}]{weidemann2000}
{Weidemann}, V. 2000, \aap, 363, 647

\bibitem[{{Weidemann} \& {Koester}(1983)}]{weidemann1983}
{Weidemann}, V., \& {Koester}, D. 1983, \aap, 121, 77

\bibitem[{{WFIRST Astrometry Working Group} {et~al.}(2019){WFIRST Astrometry Working Group}, {Sanderson}, {Bellini}, {Casertano}, {Lu}, {Melchior}, {Libralato}, {Bennett}, {Shao}, {Rhodes}, {Sohn}, {Malhotra}, {Gaudi}, {Fall}, {Nelan}, {Guhathakurta}, {Anderson}, \& {Ho}}]{WFIRST2019}
{WFIRST Astrometry Working Group}, {Sanderson}, R.~E., {Bellini}, A., {et~al.} 2019, Journal of Astronomical Telescopes, Instruments, and Systems, 5, 044005, \dodoi{10.1117/1.JATIS.5.4.044005}

\bibitem[{{Woodgate} {et~al.}(1998){Woodgate}, {Kimble}, {Bowers}, {Kraemer}, {Kaiser}, {Danks}, {Grady}, {Loiacono}, {Brumfield}, {Feinberg}, {Gull}, {Heap}, {Maran}, {Lindler}, {Hood}, {Meyer}, {Vanhouten}, {Argabright}, {Franka}, {Bybee}, {Dorn}, {Bottema}, {Woodruff}, {Michika}, {Sullivan}, {Hetlinger}, {Ludtke}, {Stocker}, {Delamere}, {Rose}, {Becker}, {Garner}, {Timothy}, {Blouke}, {Joseph}, {Hartig}, {Green}, {Jenkins}, {Linsky}, {Hutchings}, {Moos}, {Boggess}, {Roesler}, \& {Weistrop}}]{HST_STIS}
{Woodgate}, B.~E., {Kimble}, R.~A., {Bowers}, C.~W., {et~al.} 1998, \pasp, 110, 1183, \dodoi{10.1086/316243}

\bibitem[{{Yuan} {et~al.}(2013){Yuan}, {Liu}, \& {Xiang}}]{yuan2013}
{Yuan}, H.~B., {Liu}, X.~W., \& {Xiang}, M.~S. 2013, \mnras, 430, 2188, \dodoi{10.1093/mnras/stt039}

\end{thebibliography}

\appendix

\section{FORECASTOR Code Examples}\label{appdx:code_examples}

We begin by defining a \telescope instance in accordance with the \textit{CASTOR}
reference design:

\begin{minted}{python}
from castor_etc.telescope import Telescope
# Define a telescope object with default CASTOR reference design parameters
MyTelescope = Telescope()
\end{minted}

Any changes to the default parameters should be passed as arguments to the \telescope
class as opposed to modifying the source code directly. The example below demonstrates how
this is done.

\begin{minted}{python}
# Specify a new detector read noise (e/pixel):
MyNewTelescope = Telescope(read_noise=1.0)
\end{minted}

Next, we describe the sky background conditions (zodiacal light, Earthshine, and
geocoronal emission lines) via the \background class:

\begin{minted}[texcomments=true]{python}
from castor_etc.background import Background
# Use default Earthshine & zodiacal light values
MyBackground = Background()
# Add "high" [O\textit{\textsc{ii}}] 2471\AA emission line with default wavelength and linewidth
MyBackground.add_geocoronal_emission(flux="high")
# Also add Ly-\(\alpha\) airglow line
MyBackground.add_geocoronal_emission(
  wavelength=1216, # Å or `astropy.units`
  linewidth=0.04, # Å or `astropy.units`
  flux=6.1e-14, # erg/s/cm\^{}2/arcsec\^{}2
)
\end{minted}

We now explain how to generate observing targets using each of the \pointsource,
\extendedsource, and \galaxysource classes. For completeness, an example demonstrating the
use of a \customsource object is available in the \texttt{CASTOR-telescope} GitHub
organization as a Jupyter
notebook\footnote{\smash[b]{\url{https://github.com/CASTOR-telescope/ETC_notebooks/blob/master/custom_source.ipynb}}}.

\subsection{Point Sources}\label{sec:point_source}

To begin, we simulate a star as a blackbody point source of 8000\K at a redshift of
\(z\)\(=\)0.06. For blackbody spectra, we use the Planck radiation law to obtain the
spectral radiance of the blackbody in units of
\(\mathrm{erg}\pow{\s}\pow[-2]{\cm}\pow{\text{\AA}}\pow{\sr}\). To get the flux density in
units of \(\mathrm{erg}\pow{\s}\pow[-2]{\cm}\pow{\text{\AA}}\), we set this spectrum to
correspond to a star of one solar radius at a distance of 1\kpc by default (i.e., by
multiplying the spectrum by the source's projected solid angle); however, this behaviour
can be changed and its effects are usually irrelevant due to renormalization of the
spectrum. In our example, we will set the spectrum to correspond to a star of 0.5\Rsun
located 10\kpc from Earth.

\begin{minted}{python}
from castor_etc.sources import PointSource
# Create point source
MyPointSource = PointSource()
# Approximate the star as a blackbody.
# Note that only the temperature is required, but others are shown for illustration
MyPointSource.generate_bb(
  8000, # kelvin, but can use `astropy.units`
  redshift=0.06,
  limits=[900, 30000], # Å or `astropy.units`
  radius=0.5, # R$_\odot$ or `astropy.units`
  dist=10, # kpc or `astropy.units`
)
\end{minted}

\begin{figure}[htb]
  \abovecaptionskip=0pt
  \belowcaptionskip=0pt
  \centering
  \includegraphics[width=0.6\textwidth]{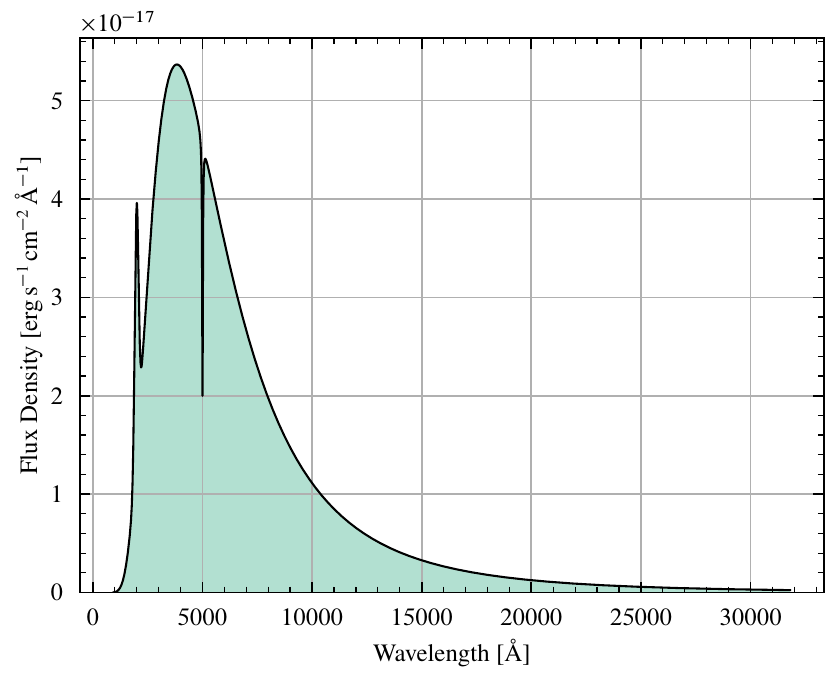}
  \caption{The generated blackbody spectrum with spectral lines visualized using
  \mintinline{python}{MyPointSource.show_spectrum()}.}
  \label{fig:pointsource_spectrum}
\end{figure}

We can add emission and absorption lines to the base spectrum as well as visualize the
source spectrum, whose output is shown in Figure \ref{fig:pointsource_spectrum}, using:

\begin{minted}{python}
# Add weirdly broad spectral lines O_o
MyPointSource.add_emission_line(
    center=2000, # Å or `astropy.units`
    fwhm=200, # Å or `astropy.units`
    peak=2.5e-17, # erg/s/cm^2/Å
    shape="gaussian",
    abs_peak=False, # add 2.5e-17 erg/s/cm^2/Å on top of continuum
)
MyPointSource.add_absorption_line(
    center=5005,
    fwhm=40,
    dip=2e-17,
    shape="lorentzian",
    abs_dip=True, # ensure line minimum is at 2e-17 erg/s/cm^2/Å
)
# Plot spectrum
MyPointSource.show_spectrum()
\end{minted}

\subsection{Extended Sources}\label{sec:extended_source}

Unlike point sources, the \extendedsource class has non-zero angular dimensions where the
flux may change with angular position. To this end, we supply two default surface
brightness profiles:~\texttt{uniform} and \texttt{exponential}. A uniform profile results
in a constant surface brightness over an elliptical region, and the surface brightness
drops to zero immediately outside this ellipse. In contrast, an exponential profile is
defined by its scale lengths along the semimajor and semiminor axes, and the surface
brightness smoothly decreases from the centre of the source out to infinity. If these
profiles are insufficient, a user can also supply a function to the \texttt{profile} class
parameter that describes some arbitrary surface brightness profile for the
\extendedsource. The following example illustrates how to generate a uniform surface
brightness profile for an extended source like a diffuse nebula.

\begin{minted}{python}
from castor_etc.sources import ExtendedSource
MyExtendedSource = ExtendedSource(
  angle_a=3, # semimajor axis, arcsec or `astropy.units`
  angle_b=1, # semiminor axis
  rotation=45, # CCW angle relative to x-axis
  profile="uniform", # can be a function
)
\end{minted}

We now assign a Gaussian emission line spectrum to this source, and renormalize the
spectrum so it has an AB magnitude of 25 in the u-band. Under the hood, we use Simpson's
rule to numerically integrate Eq.~(2) of \citet{Bessell2012}, interpolating the passband
response curves to the wavelength resolution of the spectrum if necessary. We interpolate
to the spectrum resolution for two main reasons. First, in most cases, the spectrum is
higher resolution than our bandpass response curves. Second, the curves in our bandpass
files are relatively well-behaved compared to observational spectra, which may have lots
of sharp peaks and troughs. If we interpolate the spectrum to the bandpass resolution, we
risk losing these features that may have a substantial contribution to our calculations.
In contrast, since the passband throughput curves are smoother, any interpolation (even to
a coarser spectrum) should capture the behaviour of the passband reasonably well.

\begin{minted}{python}
# Emission line spectrum
MyExtendedSource.generate_emission_line(
  center=2500, # Å or `astropy.units`
  fwhm=2, # Å or `astropy.units`
  tot_flux=1e-19, # the total flux under the curve.
                  # Can alternatively specify the peak of the emission line
  shape="gaussian", # "gaussian" or "lorentzian"
)
# Renormalize for illustrative purposes
MyExtendedSource.norm_to_AB_mag(
  25, # AB magnitude
  passband="u",
  TelescopeObj=MyTelescope,
)
\end{minted}

Note that we take the response function to be unity when renormalizing to a specific
\emph{bolometric} AB magnitude. In the notation of
\citet{Bessell2012}:~\(S_x(\lambda)=1\). In this case, it is also important to ensure that
the spectrum used in bolometric AB magnitude calculations is sufficiently small at the
edges. If the spectrum does not vanish at the ends, like a uniform spectrum, then the
bolometric magnitude will depend on the length of the spectrum because the area under the
curve does not converge.

\subsection{Galaxy Sources}\label{sec:galaxy_source}

Finally, the \galaxysource class is similar to \extendedsource, except the surface
brightness profile follows a S\'ersic model. The user supplies \galaxysource with an
effective (half-light) radius, S\'ersic index, and axial ratio---the ratio of semiminor to
semimajor axis---which the code uses to generate its surface brightness profile, e.g.,

\begin{minted}{python}
from castor_etc.sources import GalaxySource
MyGalaxySource = GalaxySource(
    r_eff=3, # arcsec or `astropy.units`
    n=4, # Sérsic index
    axial_ratio=0.9, # semiminor/semimajor axis
    rotation=135, # CCW rotation from x-axis
)
\end{minted}

Next, we will use one of the galaxy spectra available in \cetc and renormalize it to
correspond to a given luminosity and distance.

\begin{minted}{python}
MyGalaxySource.use_galaxy_spectrum("spiral")
MyGalaxySource.norm_luminosity_dist(
  luminosity=2.6e10, # L$_\odot$ or `astropy.units`
  dist=765 # kpc or `astropy.units`
)
\end{minted}

The luminosity-distance normalization is accomplished by dividing each spectrum value by
the total luminosity of the spectrum, and then multiplying by the desired luminosity. We
use Simpson's rule to integrate the spectrum, in units of
\(\erg\s^{-1}\cm^{-2}\text{\AA}^{-1}\), to get the total radiance of the object in units
of \(\erg\s^{-1}\cm^{-2}\). We assume the object emits radiation isotropically from a
distance \(d\), meaning the total luminosity of the object is simply the total radiance
multiplied by \(4\pi d^2\). This total luminosity is the denominator within our
normalization factor that we multiply with the spectrum, with the desired luminosity as
the numerator.

\subsection{Photometry Calculations}

To execute photometry calculations, we initialize a \photometry object with our
\telescope, \source, and \background instances. Below, we select an ``optimal'' aperture
for our point source.

\begin{minted}{python}
from castor_etc.photometry import Photometry
# Create photometry object
MyPointPhot = Photometry(MyTelescope, MyPointSource, MyBackground)
# Specify the aperture with an optional factor
MyPointPhot.use_optimal_aperture(factor=1.35)
# Aperture width = factor × telescope's FWHM
\end{minted}

If the user is not doing point source photometry, or prefers to use a different aperture,
then we can specify a rectangular or elliptical aperture for the photometry calculations
via:

\begin{minted}{python}
import astropy.units as u # for convenience

MyExtendedPhot = Photometry(MyTelescope, MyExtendedSource, MyBackground)
# Specify off-centre rectangular aperture
MyExtendedPhot.use_rectangular_aperture(
  width=4.5 * u.arcsec, length=3 * u.arcsec, center=[0.5, -1] * u.arcsec
)

MyGalaxyPhot = Photometry(MyTelescope, MyGalaxySource, MyBackground)
# Specify centred elliptical aperture
MyGalaxyPhot.use_elliptical_aperture(
    a=6 * u.arcsec,
    b=4 * u.arcsec,
    center=[0, 0] * u.arcsec,
    rotation=31.41592654, # degree
)
\end{minted}

\begin{figure*}[htb]
  \centering
  \begin{minipage}[c]{0.32\textwidth}
    \vspace*{0pt}
    \includegraphics[width=\textwidth]{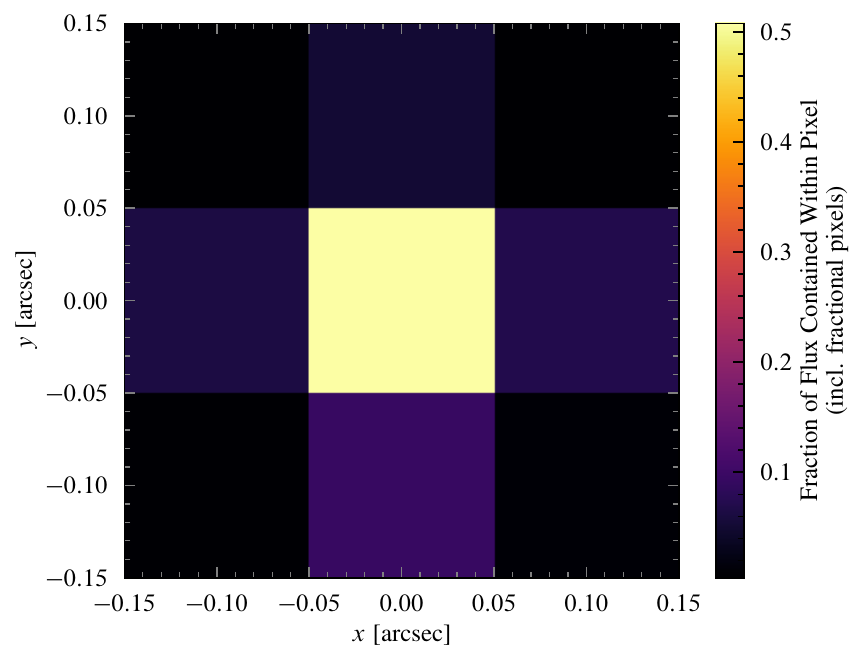}
  \end{minipage}
  \hfill
  \begin{minipage}[c]{0.32\textwidth}
    \vspace*{0pt}
    \includegraphics[width=\textwidth]{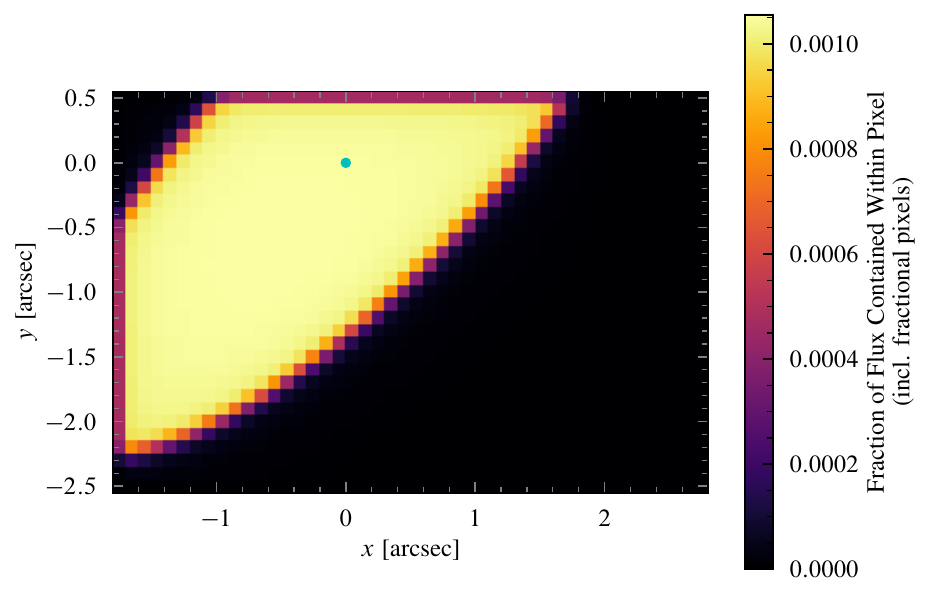}
  \end{minipage}
  \hfill
  \begin{minipage}[c]{0.32\textwidth}
    \vspace*{0pt}
    \includegraphics[width=\textwidth]{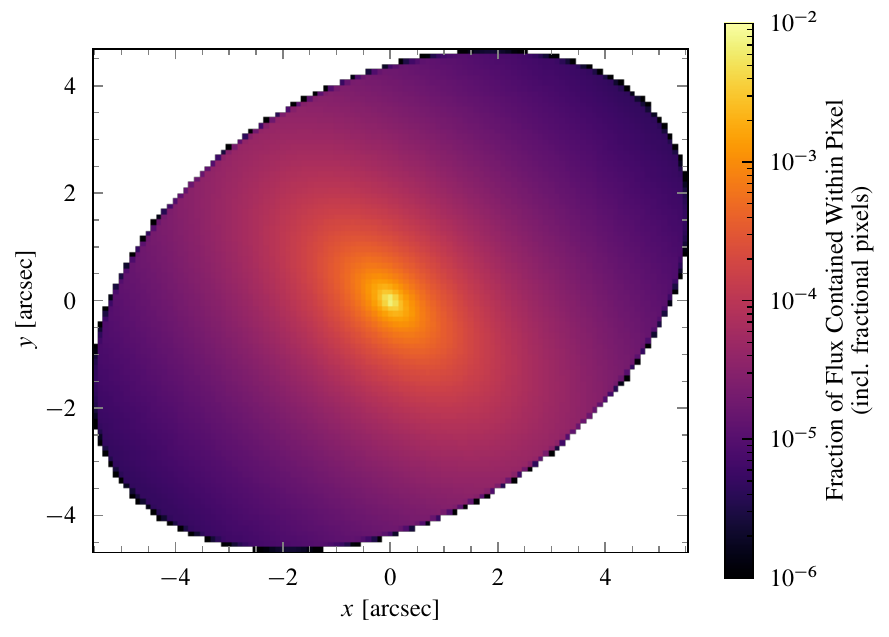}
  \end{minipage}
  \caption{A point source (left), extended source (centre), and galaxy (right) as seen
   through an optimal circular aperture (left), rectangular aperture (centre), and
   elliptical aperture (right) in the g-band. From left to right, the apertures enclose
   80.60\%, 58.51\%, and 59.01\% of the flux from these sources.}
  \label{fig:apertures}
\end{figure*}

FORECASTOR ETC has a built-in method to visualize these sources through their apertures in
a given passband, which we invoke with the following code. The plots generated by the ETC
are shown in Figure \ref{fig:apertures}.

\begin{minted}[mathescape]{python}
# Code to produce plots in Figure $\it \ref{fig:apertures}$
from matplotlib.colors import LogNorm

MyPointPhot.show_source_weights("g")
MyExtendedPhot.show_source_weights("g", mark_source=True)
MyGalaxyPhot.show_source_weights("g", norm=LogNorm(vmin=1e-6, vmax=0.01))
\end{minted}

Finally, we can use the following two methods to calculate the \sn achieved over a certain
integration time and to calculate the integration time needed to achieve a given \sn. We
can optionally supply a value for the reddening associated with the source and the
telescope's pointing. The results of these calculations are tabulated in Table
\ref{tab:snr_times}, along with the AB magnitudes through \textit{CASTOR}'s passbands
obtained through \mintinline{python}{MyPointSource.get_AB_mag(MyTelescope)} and the
encircled energy in each passband.

\begin{minted}[mathescape]{python}
# Results are tabulated in Table $\it \ref{tab:snr_times}$
time_to_achieve_snr = MyPointPhot.calc_snr_or_t(
  snr=10,
  reddening=0.01, # E(B-V), optional
)
snr_obtained_with_t = MyPointPhot.calc_snr_or_t(
  t=314.15,
  reddening=0.01, # E(B-V), optional
)
\end{minted}

\begin{deluxetable}{ccccc}[H]
  \tablecaption{Results of sample \pointsource \sn calculations and AB
  magnitudes.\label{tab:snr_times}}
  \tablehead{
    \multirow{2}{*}{Passband} &
    \multirow{2}{*}{AB Mag} &
    Encircled &
    Time (s) to &
    \sn After
    \\[-2pt]
    &
    &
    Energy (\%) &
    \(\sn=10\) &
    \(t=314.15\s\)
  }
  \decimals
  \startdata
  UV & 22.12 & 91.36 &            17.08 & \hphantom{0}48.14 \\
  u  & 20.62 & 81.07 & \hphantom{0}4.37 & \hphantom{0}95.72 \\
  g  & 20.04 & 80.60 & \hphantom{0}2.08 &            138.97 \\
  \enddata
\end{deluxetable}

\section{Adapting FORECASTOR to Other Missions}\label{appdx:adapting_forecastor}

The main step in adapting FORECASTOR ETC to other missions is to create a \telescope
instance that represents the physical telescope used in the mission. All other classes and
methods are either agnostic to the telescope or directly reference the parameters of the
given \telescope instance; we do not assume any \textit{CASTOR}-specific values in any
calculations outside of the default \telescope parameters.

The list of default \telescope parameters is contained in \cetc's \texttt{parameters.py}
file. The user should \emph{not} modify this file directly; instead, all customizations
can be accomplished when instantiating a new \telescope object by passing keyword
arguments. The \texttt{parameters.py} file should only serve as a tool for determining
which parameters need to be modified.

Following is an example showing how to create a new \telescope instance with all
parameters relevant for the photometry calculations explicitly listed. Certain arguments
may have specific requirements (e.g., the format of the passband response curves), and
these are documented in the \telescope object's docstring.

\begin{minted}[mathescape]{python}
import astropy.units as u
from castor_etc.telescope import Telescope

# Define the parameters to customize
custom_params = {
  # The name of the passbands
  "passbands": ["my_passband1", "my_passband2", "my_passband3"],
  # The [lower, upper] wavelength cutoffs for the passbands
  "passband_limits": {
    "my_passband1": [123, 321] * u.nm,
    "my_passband2": [456, 654] * u.nm,
    "my_passband3": [789, 987] * u.nm,
  },
  # The files containing the passband response curves.
  # These need to be plain text files. See docstring for more details
  "passband_response_filepaths": {
    "my_passband1": "my_passband_curve1.txt",
    "my_passband2": "my_passband_curve2.txt",
    "my_passband3": "my_passband_curve3.txt",
  },
  # The units of the wavelength columns in the passband response files
  "passband_response_fileunits": {"my_passband1": u.nm, "my_passband2": u.nm, "my_passband3": u.nm},
  # The desired linear interpolation resolution of the passband response curves.
  # If None, use the native resolution of the passband response curves
  "passband_resolution" 1 * u.AA,  # highly recommended to set to not None
  # Keyword arguments for finding the photometric zero-points. See docstring for full details
  "phot_zpts_kwargs": {
    "method": "secant", # "secant" or "bisection"
    # The two initial guesses for the secant method or the bounds for the bisection method
    "ab_mags": {
      "my_passband1": [25.5, 23.5], "my_passband2": [25.5, 23.5], "my_passband1": [25.5, 23.5]
    },
    "tol": 2e-4, # the desired accuracy of the photometric zero-points
    "max_iter": 100, # maximum number of iterations to use for finding each zero-point
  },
  # The filepaths to each passband's PSF
  "psf_filepaths": {
    "my_passband1": "my_passband_PSF1.fits",
    "my_passband2": "my_passband_PSF2.fits",
    "my_passband3": "my_passband_PSF3.fits",
  },
  # The PSF oversampling factor
  "psf_supersample_factor": 20,
  # The full-width at half-maximum of the PSF. Only used for estimating the "optimal aperture" size
  "fwhm": 0.15 * u.arcsec,
  # The linear angle subtended by each square pixel in the detector
  "px_scale": 0.1 * u.arcsec,
  # The dark current in units of electrons/s per pixel
  "dark_current": 1e-4
  # The read noise in units of electrons/pixel
  "read_noise": 3.0
  # The maximum wavelength beyond which we consider the flux to be red leak for the given passband
  "redleak_thresholds": {
    "my_passband1": 3000 * u.AA, "my_passband2": 6000 * u.AA, "my_passband3": 9000 * u.AA
  }
  # The extinction coefficients (i.e., R := A/E(B-V)) for each passband
  "extinction_coeffs": {"my_passband1": 9.42, "my_passband2": 6.28, "my_passband3": 3.14}
}

# Create a `Telescope` instance with these custom parameters
MyCustomTelescope = Telescope(**custom_params)
\end{minted}

After defining a \telescope instance suitable for the mission, all other functionality of
the ETC remains unchanged. This makes simulating different telescope parameter
combinations a straightforward task, as only one piece of code needs to be updated while
the rest can be reused.
\end{document}